\documentclass{aa}
\usepackage[varg]{txfonts}

\usepackage[normalem]{ulem}

\usepackage[switch]{lineno}

\usepackage[utf8]{inputenc}
\usepackage[usenames,dvipsnames]{color}
\usepackage[caption=false]{subfig}
\usepackage{graphicx,url,twoopt,natbib}
\usepackage{tikz} 
\usetikzlibrary{shapes,arrows} 

\usepackage[breaklinks=true]{hyperref} 
\makeatletter

\bibpunct{(}{)}{;}{a}{}{,}    
\makeatletter
  \protected\def\stonyslink{%
     \def\hyper@linkstart##1##2{}\let\hyper@linkend\@empty}
  \newcommandtwoopt{\citeads}[3][][]{%
   \href{http://ui.adsabs.harvard.edu/abs/#3/abstract}%
        {\stonyslink \citealp[#1][#2]{#3}}
   \biblink{#3}{\href{http://ui.adsabs.harvard.edu/abs/#3/abstract}{ADS}}}
 \newcommandtwoopt{\citepads}[3][][]{%
   \href{http://ui.adsabs.harvard.edu/abs/#3/abstract}%
        {\stonyslink \citep[#1][#2]{#3}}
   \biblink{#3}{\href{http://ui.adsabs.harvard.edu/abs/#3/abstract}{ADS}}}
 \newcommandtwoopt{\citetads}[3][][]{%
   \href{http://ui.adsabs.harvard.edu/abs/#3/abstract}%
        {\stonyslink \citet[#1][#2]{#3}}
  \biblink{#3}{\href{http://ui.adsabs.harvard.edu/abs/#3/abstract}{ADS}}}
 \newcommandtwoopt{\citeyearads}[3][][]{%
   \href{http://ui.adsabs.harvard.edu/abs/#3/abstract}%
        {\stonyslink \citeyear[#1][#2]{#3}}
   \biblink{#3}{\href{http://ui.adsabs.harvard.edu/abs/#3/abstract}{ADS}}}
\makeatother

\begin{document}

\title{The optical-infrared relation for active galactic nuclei: The role of contaminations}

\author{Mateusz Ra{\l}owski\inst{1,}\inst{3,}\inst{4} 
    \and Krzysztof Hryniewicz\inst{2}
    \and Katarzyna Ma{\l}ek\inst{2}
  \and Agnieszka Pollo\inst{1,}\inst{2}
    \and Guido Risaliti\inst{4,}\inst{5}}

\offprints{M. Ra{\l}owski, \email{mralowski@oa.uj.edu.pl}}

\institute{Astronomical Observatory of the Jagiellonian University, Faculty of Physics, Astronomy and Applied Computer Science, ul. Orla 171, 30-244 Cracow, Poland; \email{mralowski@oa.uj.edu.pl}
  \and National Centre for Nuclear Research, ul. Pasteura 7, 02-093 Warsaw, Poland;
  \and Jagiellonian University, Doctoral School of Exact and Natural Sciences, Astronomy
  \and Dipartimento di Fisica e Astronomia, Università di Firenze, via G. Sansone 1, 50019 Sesto Fiorentino, Firenze, Italy
  \and INAF-Osservatorio Astrofisico di Arcetri, Largo E. Fermi 5, 50125, Firenze, Italy}

\date{Received ***** / Accepted 15.01.2026}

\abstract 
{Although the population of quasars evolved significantly in the past, the properties of quasars as physical objects are supposed to remain almost unchanged, which makes quasars promising candidates for cosmological tests. The X–UV luminosity relation in particular is widely used for this purpose. However, the potential of other spectral domains for this purpose remains open.}
{The aim of the analysis we present is to test the parameter space in order to build a well behaving OPT-IR correlation that could serve as a cosmological probe. The main objective is to calibrate the OPT-IR luminosity relation for quasars, focusing on accurate estimations of dusty torus and accretion disk luminosities. We analyzed contaminations related to host galaxies, particularly from polar dust, the interstellar medium, and stellar emission that affect the optical and infrared.}
{We used a sample of nearly 400 quasars with photometrical observations and spectroscopical redshift divided into four redshift bins (0.7–2.4). Full spectral energy distribution (SED) fitting was performed with the CIGALE code, and results were compared with simplified photometric luminosity estimates. The impact of non-active galactic nucleus components and the role of polar dust in the fitting process were assessed.}
{We show that for sources with a disk luminosity above 10$^{45}$ [erg s$^{-1}$], the photometric estimates are consistent with SED-based values. While polar dust contributes marginally to luminosity, its presence significantly alters SED fitting, particularly the torus opening angle and cold dust properties. In the optical domain, stellar emission is the dominant contamination. In the infrared, disk emission and cold dust play major roles. We propose two empirical calibrations for the OPT–IR relation.}
{We conclude that the optical band is dominated by the accretion disk component above 10$^{45}$ or 10$^{46}$ [erg s$^{-1}$] depending on redshift, while IR luminosity is dominated by the dusty torus emission above $1.6\times10^{45}$ or $2\times10^{46}$ [erg s$^{-1}$] depending on the redshift. In this high-luminosity regime, simplified photometric methods yield reliable disk and torus luminosity estimates.}

\keywords{galaxies: active - quasars: general – galaxies: statistics}
\maketitle

\section{Introduction}

The unification schemes for active galactic nuclei (AGNs; \citealt{antonucci1985, urry1995, Netzer_2015, Hickox2018}) were proposed to provide a simplified explanation of the structure of the AGN mechanism. Unification based on the inclination angle, where the dusty torus hides some of the internal structures, was questioned with superior observations with Gravity \cite{Gravity2020}. Recent studies suggest that the structure of the torus might be more complex (consisting of clumps and clouds) \cite{Nenkova2008} and that it can be directly linked to the broad-line regions (BLRs) and the accretion disk \citep{Czerny2011, Czerny2019MgII}. There have also been indications that there is dust located in the polar regions of AGNs, so-called polar dust (PD; \citealt{Toba_2021, Hide_dust_2022, Dust2022MNRAS}). The dusty torus can change the obscuration along the line of sight, its spectral properties, and the classification of an AGN. Additionally, some authors have found a significant variance in the amount of dust in AGNs (e.g., \citealt{CF_var_2013}).

Among AGN, quasars are particularly important due to their extreme brightness and wide redshift distribution. In recent years, significant progress has been made in calibrating quasars as standard candles for cosmological tests. The search for new standard candles is crucial, especially considering the pressing issue in modern observational cosmology known as the "Hubble tension" \citep{Planck2014, Verde2013, Planck2020, Verde2019, Riess2022}. Recent studies have suggested that quasars observed with JWST at a high redshift ($z>2$) may exhibit different properties or physical structures compared to local quasars \citep{Maiolino2024}.

Previous approaches utilized quasars for cosmology by making use of two major methodologies. The first methodology is based on the technique known as reverberation mapping \citep{Kaspi2000, Bentz2013, Czerny2011, Czerny2013, Zajacek2020,  Czerny2019MgII, Czerny2021}. This technique involves calculating the time lag between emissions from different regions within the AGN — specifically, between the accretion disk surrounding the supermassive black hole at the quasar's center and the BLR around it. From the time lag, the radius of more distant regions can be calculated, along with the absolute luminosity. Using the absolute luminosity, the luminosity distance can be determined, and by comparing it to the redshift, cosmological models can be tested. The primary drawback of this method is its time consumption, as measuring the time lag requires years of observations for each object. Additionally the high dispersion of the R(BLR)-L method makes it problematic to perform reverberation mapping outside of the local Universe.

An alternative approach is based on the nonlinear luminosity relation proposed in the Risaliti-Lusso relation \citep{RisalitiLusso2015, Risaliti2019Nat, Signorini2023}. The Risaliti-Lusso relation is based on a widely accepted scheme that the ultraviolet emission of AGN is produced by an accretion disk, while the X-rays are produced by a hot corona around the accretion disk. Studies \citep{Tananbaum1979, Zamorani1981, Avni1986} have shown that the relationship between X-ray and UV luminosities can be parametrized as $\log(L_{\text{X}})$ = $\gamma\log(L_{\text{UV}})$ + $\beta$, where $L_{\text{X}}$ and $L_{\text{UV}}$ are the rest-frame monochromatic luminosities at 2 keV and 2,500 \AA, respectively. Although $L_{\text{X}}$ and $L_{\text{UV}}$ are measured independently, they are physically correlated with each other and with the luminosity distance.

The optical (OPT) and infrared (IR) luminosity relation for quasars also shows promise. Several studies have indicated that this relation can be robust \citep{2013ApJ...773..176G, Ralowski2024, Trefoloni2024}. The reason is that IR luminosity (L$_{IR}$) and optical luminosity (L$_{OPT}$) should be strictly connected. In bright quasars, the bulk of L$_{OPT}$ emission can be explained by the emission from the accretion disk \cite{Netzer_2015}. Part of this emission is absorbed and reemited by the dusty torus surrounding the supermassive black hole. Thus L$_{OPT}$ and L$_{IR}$ should be tightly correlated. The degree of this correlation is still not clear, but if it were possible to account for 1) potential selection effects and 2) any physical contaminations, it may then be possible to use it as a universal relation that is constant with redshift. In a previous study, \cite{Ralowski2024} we showed that the OPT–IR relation may be influenced by selection effects in the data, particularly involving the W3 and W4 filters in the IR, and that the covering factor, calculated from the $L_{IR}$/$L_{agn}$ luminosity ratio, does not evolve with redshift for quasars with similar physical properties. Our other conclusion was that the OPT-IR relation has to be calibrated, as it is influence by contamination, such as from polar dust.

The importance of such a calibration is even more significant because we believe that the OPT-IR nonlinear luminosity relation can be used for cosmological tests. In essence, it is similar to the X-UV Risaliti-Lusso relation that is widely used and debated today. For the OPT-IR photometrical relation to be useful for cosmology, it is necessary to exclude or/and understand the 1) systematics and the 2) additional contributions to the OPT or IR emission from components of the quasar host galaxy that are not connected to AGN emission. In the following, we analyze the photometrical observations, especially through the spectral energy distribution (SED) fitting, to find the most important factors and contaminations that can change the OPT-IR relation connected only to the central source AGN.

To calibrate the luminosity relation for objects as complex as quasars, dependencies on physical parameters must be carefully analyzed. Some works (e.g., \citealt{Toba_2021, Buat2021}) have indicated that the L$_{IR}$-L$_{opt}$ relation can depend on physical properties of AGN such as polar dust. It is also possible that the stellar light of the host galaxy influences the L$_{OPT}$. In \cite{Ralowski2024} we discuss possible explanations of how the low-luminosity part of the low-$z$ sample breaks the trend set by the high-mass part in L$_{IR}$/L$_{OPT}$. This is in line with the work by \cite{Toba_2021}, who suggested strong polar dust contamination in L$_{IR}$ in the lower-luminosity part of the AGN population. Reddening and dust obscuration at small redshift was reported to affect type-II AGNs more than type-I AGNs \citep{Netzer_2015}. The solution to this may be selecting the samples based on the IR, but this approach would be biased toward sources with a larger covering factor and against sources with large inclination angles (see Sect. 2 in \citealt{Netzer_2015}). Another important contamination may be emission from star-forming and inter regions in the galaxy as well as interstellar medium (ISM). The ISM also has an important effect on high redshift quasars \citep{Decarli2023} and line emission \citep{Kade2024}.

With the work presented here, we aim to analyze the impact of such contaminations while focusing on two key aspects: 1) the role of contaminations in luminosity estimation and how they affect the OPT–IR relation, and 2) whether it is possible to calibrate our method after accounting for these contaminations. The following work is structured as follows: Section \ref{sec:Data} presents the data selection process, in Sect. \ref{sec:methods} we introduce the methodology we used, and Sect. \ref{sec:results} presents the most important results.

\section{Data}
\label{sec:Data}
The primary dataset used in this analysis is based on photometric observations of quasars. Specifically, the Sloan Digital Sky Survey Quasar Catalog, Sixteenth Data Release (SDSS DR16Q\footnote{\url{https://data.sdss.org/datamodel/files/BOSS_QSO/DR16Q/DR16Q_v4.html}}; \citealt{2020ApJS..250....8L}), was employed. This catalog contains approximately 750,000 quasars, including around 225,000 newly identified sources relative to previous data releases. Quasars in SDSS DR16Q were classified using spectral fitting applied to the SDSS-IV/Extended Baryon Oscillation Spectroscopic Survey (eBOSS) spectra \citep{Dawson2016eBOSS}. The final data catalog is estimated to have a completeness rate of 99.8\%, with defilement levels ranging from 0.3 to 1.3\% (for further details, see \citealt{2020ApJS..250....8L}). The SDSS photometric data are provided in two formats: "asinh magnitudes" and "nanomaggies"\footnote{\url{https://www.sdss.org/dr16/algorithms/magnitudes/\#Fluxunits:maggiesandnanomaggies}} \citep{1999AJ....118.1406L}. The fluxes in nanomaggies can be converted to AB magnitudes after applying corrections for filter-specific shifts in the \textit{u} and \textit{z} filters.\footnote{\url{https://www.sdss.org/dr12/algorithms/fluxcal/\#SDSStoAB}}

There are 557,000 objects after the redshift selection within the range of z=0.7-2.4.
SDSS DR16Q was subsequently crossmatched using a 3 arcsecond radius. After adding the IR from the Wide-field Infrared Survey Explorer (WISE; \citealt{wright2010}) All-Sky Survey, the sample had 552,372 objects. Near-IR photometry was obtained from The UKIRT Infrared Deep Sky Survey (UKIDSS, \citealt{2007MNRAS.379.1599L}) resulting in just over 100,000 objects in the sample. For the ultraviolet wavelength range, we used the photometry from the Galaxy Evolution Explorer (GALEX GR6/GR7 data release \citep{martin2005}, which we restricted to low-redshift quasars, as the Lyman-$\alpha$ break becomes significant at higher redshifts. The UV observations were available for 17,463 objects with redshift z$\leq$1.1. Far-IR (FIR) data were gathered from the Herschel mission, specifically from the Photodetector Array Camera and Spectrometer (PACS; \citealt{PACS2010}) and Spectral and Photometric Imaging Receiver (SPIRE; \citealt{SPIRE2010}) instruments. For the Hershel data, we used the IRSA database, with the main condition being for the quasars to have SPIRE 250$\mu$m. The inclusion of Herschel data shrank the number of the whole sample massively, leaving us with only 351 objects. Additionally, SPITZER \citep{Spitzer_catalog2021} data were available for a few objects. In the procedure, we also included upper limits, which were accounted for in the SED fitting procedure (described in Sect. \ref{sec:SED}). SDSS, UKIDSS, WISE, and 250$\mu$m SPIRE all have 100\% coverage for all the selected objects. The data binning with respect to the redshift is presented below. For each dataset in brackets, the fraction of coverage in the remaining surveys is presented.

We split SDSS DR16Q into four samples based on the redshift ranges (see Fig. \ref{fig:chart}): 
 \begin{enumerate}
     \item Low-$z$: z $\in$ [0.7, 1.1], containing 100 objects 
    
     (PACS: blue 0\%, green 2\%, red 4\%; SPIRE: 350$\mu$m 30\%, 500$\mu$m 6\%)
     
     \item Inter-$z1$: z $\in$ (1.1, 1.5), containing  73 objects 
     
      (PACS: blue 0\%, green 1\%, red 0\%; SPIRE: 350$\mu$m 22\%, 500$\mu$m 1\%)
     
     \item Inter-$z2$: z $\in$ [1.5, 2.0), containing 94 objects
     
      (PACS: blue 0\%, green 5\%, red 3\%; SPIRE: 350$\mu$m 34\%, 500$\mu$m 4\%)
     
     \item High-$z$: z $\in$ [2.0, 2.4], containing 84 objects 
     
      (PACS: blue 1\%, green 0\%, red 1\%; SPIRE: 350$\mu$m 37\%, 500$\mu$m 6\%)
 \end{enumerate}
 
 The observations were corrected for extinction using the Cardelli extinction law \citep{1989ApJ...345..245C} and dust maps from the work of \citep{1998ApJ...500..525S}.

\tikzstyle{decision} = [rectangle, draw, fill=green!20, text width=4.5em, text badly centered, node distance=2cm, inner sep=5pt]
\tikzstyle{block} = [rectangle, draw, fill=teal!40, node distance=1.8cm , text width=5em, text centered, rounded corners, minimum height=4em]
\tikzstyle{block1} = [rectangle, draw, fill=magenta!20, node distance=1.8cm , text width=4.8em, text centered, rounded corners, minimum height=4em]

\tikzstyle{line} = [draw, -latex']

\tikzstyle{cloud_low_high} = [draw, rectangle, fill=gray!30, node distance=2.1cm, text width=4.5em, text centered, rounded corners, minimum height=4.em]

\tikzstyle{cloud_low_spitzer} = [draw, rectangle, fill=orange!50, node distance=2.1cm, text width=5em, text centered, rounded corners, minimum height=4.em]
\tikzstyle{cloud_high_spitzer} = [draw, rectangle, fill=red!50, node distance=2.1cm, text width=5em, text centered, rounded corners, minimum height=4.em]

\tikzstyle{cloud_low} = [draw, rectangle, fill=cyan!40, node distance=2.1cm, text width=5.8em, text centered, rounded corners, minimum height=5em]
\tikzstyle{cloud_i1} = [draw, rectangle, fill=orange!40, node distance=2.1cm, text width=5.8em, text centered, rounded corners, minimum height=5em]
\tikzstyle{cloud_i2} = [draw, rectangle, fill=green!40, node distance=2.1cm, text width=5.8em, text centered, rounded corners, minimum height=5em]
\tikzstyle{cloud_high} = [draw, rectangle, fill=red!40, node distance=2.1cm, text width=5.8em, text centered, rounded corners, minimum height=5em]

\begin{figure}[!htp]
\begin{tikzpicture}[node distance = 1cm, auto]
\node [block, node distance=2cm] (SDSS) {SDSS DR16Q};

\node [block, below of=SDSS, node distance=2.cm] (WISE) {WISE};
\node [block, left of=WISE, node distance=2.3cm] (UKIDSS) {UKIDSS};
\node [block, below of=UKIDSS, node distance=2.cm] (GALEX) {GALEX\\};
\node [block, right of=WISE, node distance=2.3cm] (SPITZER) {SPITZER};
\node [block, right of=SPITZER, node distance=2.3cm] (HERSCHEL) {HERSCHEL};

\node [block1, below of=GALEX, node distance=2.cm] (Low) {Redshift $z$=[0.7,1.1]};
\node [block1, right of=Low, node distance=2.3cm] (Inter1) {Redshift $z$=(1.1,1.5)};
\node [block1, right of=Inter1, node distance=2.3cm] (Inter2) {Redshift $z$=[1.5,2)};
\node [block1, right of=Inter2, node distance=2.3cm] (High) {Redshift $z$=[2,2.4]};

\node [cloud_low, below of=Low, node distance=3cm] (Low_nb) {\textbf{Low-$\mathbf{z}$}\\ 100 \\
$\chi^2$<2.5\\
\textbf{83}\\
L$_{disk}$>10$^{45}$\\
\textbf{42}\\ 
L$_{disk}$<10$^{45}$\\  41};

\node [cloud_i1, below of=Inter1, node distance=3cm] (Inter1_nb) {\textbf{Inter-$\mathbf{z1}$}\\ 73 \\
$\chi^2$<2.5\\
\textbf{60}\\
L$_{disk}$>10$^{45}$\\
\textbf{43}\\
L$_{disk}$<10$^{45}$\\  17};

\node [cloud_i2, below of=Inter2, node distance=3cm] (Inter2_nb) {\textbf{Inter-$\mathbf{z2}$}\\ 94 \\ 
$\chi^2$<2.5\\
\textbf{58}\\
L$_{disk}$>10$^{45}$\\
\textbf{50}\\
L$_{disk}$<10$^{45}$\\  8};

\node [cloud_high, below of=High, node distance=3cm] (High_nb) {\textbf{High-$\mathbf{z}$}\\ 74 \\ 
$\chi^2$<2.5\\
\textbf{60}\\
L$_{disk}$>10$^{45}$\\
\textbf{36}\\
L$_{disk}$<10$^{45}$\\  24};

\path [line, dashed] (SDSS) -- (WISE);
\path [line, dashed] (UKIDSS) -- (WISE);
\path [line, dashed] (HERSCHEL) -- (SPITZER);
\path [line, dashed] (SPITZER) -- (WISE);

\path [line, dashed] (WISE) -- (Inter1);
\path [line, dashed] (GALEX) -- (Low);
\path [line, dashed] (WISE) -- (Inter2);
\path [line, dashed] (WISE) -- (High);
\path [line, dashed] (WISE) -- (GALEX);

\path [line] (Low) -- (Low_nb);
\path [line] (Inter1) -- (Inter1_nb);
\path [line] (Inter2) -- (Inter2_nb);
\path [line] (High) -- (High_nb);

\end{tikzpicture}
\caption{Flowchart showing the data selection process as described in Sect. \ref{sec:Data}. The $\chi^2$ values were calculated for the best-fit SED in CIGALE. The L$_{disk}$ values are from the best-fit estimation of the SKIRTOR model, as described in Sect. \ref{sec:estimates_lum}.}
\label{fig:chart}

\end{figure}
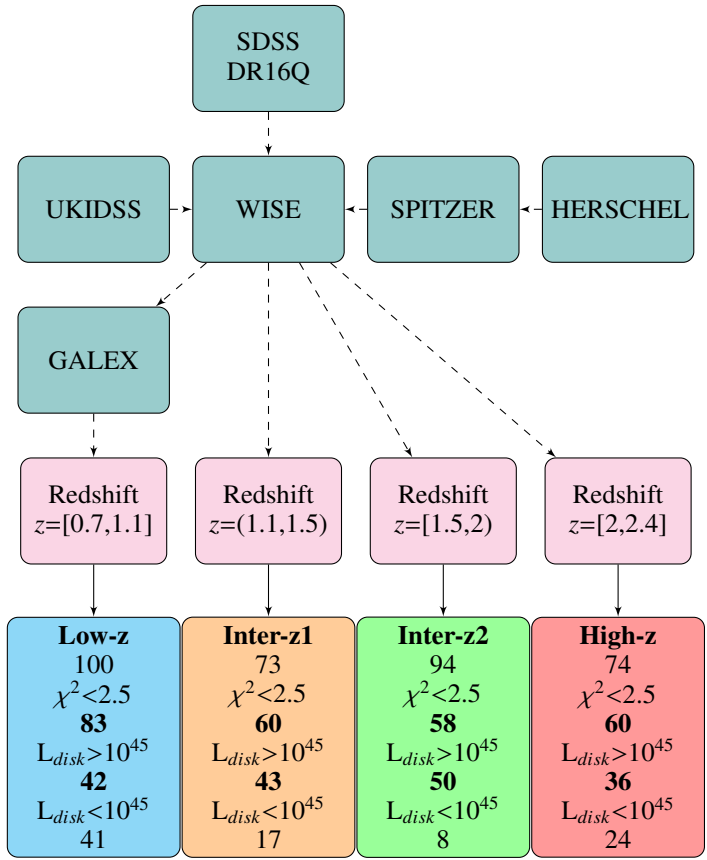

\section{Methods}
\label{sec:methods}

\subsection{SED Fitting}
\label{sec:SED}

To properly assess the physical properties of a galaxy, it is important to carefully analyze its emitted spectrum. The SED fitting technique provides insight into the broad energy distribution of photons across the electromagnetic spectrum. Using broad-band photometry that spans a wide wavelength range combined with redshift information, we constructed an observed SED. Subsequently, synthetic SEDs were generated from a set of models based on parameter-grid templates, allowing for the estimation of the underlying physical properties

The SED fitting was conducted using the Code Investigating GALaxy Emission (CIGALE; \citealt{Cigale2019}), version 2022.1.\footnote{\url{https://cigale.lam.fr/}} The CIGALE methodology is based on the assumption of an energy balance principle, and thus the energy emitted by dust in the mid-IR (MIR) to FIR was estimated from the energy absorbed by dust in the UV-optical range. The dust emission can be even larger when taking the presence of the AGN and its dusty torus into account. This approach combines computational speed and a wide range of possible applications to galaxies \citep{Buat2011, Buat2018, Boquien2013, Lo_Faro2017, Salim2018, Malek2018}, AGNs \citep{Ciesla_2015, Toba_2021}, 
ultraluminous IR galaxies (\citealt{Malek2017, Dey2024}). An important aspect of CIGALE is the use of the Bayesian-like approach. In addition, the physical properties are estimated by weighing all the models on the basis of their goodness of fit, with the best-fit models having the heaviest weight. (For more details we refer to \cite{Noll2009, Cigale2019}.)

The CIGALE modules include star formation history (SFH), stellar population synthesis, a dust attenuation model, cold dust emission, nebular emission, AGN emission, and X-ray and radio emission (these last two modules were not used). The SFH module calculates the evolution of different stellar populations based on different SFH models, taking into account different ages of stars and the $\tau$-folding time. From these calculations, using the library of stellar population synthesis models, the spectra were generated. For the SFH model, we decided to use the one already available in the CIGALE code with delayed exponential burst, which can model both starburst and older stellar populations. For the population synthesis module, we used the Bruzual and Charlot stellar library \citep{BC2003}, with the Salpeter \citep{Salpeter1951} initial mass function. Dust attenuation and cold dust models estimate the fraction of the cold dust in the host galaxy, its absorption of UV light, and later its thermal emission in IR. There is a discussion on which dust attenuation model is best for different sources of objects \citep{Malek2018}. We decided to use the Charlot and Fall (CF00, \citealt{CF2000}) attenuation law, which takes into account the dust clouds and HII regions around young stars, as it allowed us to precisely model each component.

 To model cold dust emission, we chose the model proposed in \cite{Dale2014}, which includes the polycyclic aromatic hydrocarbon and the radiation field. Thanks to the inclusion of these elements, the model is very flexible for the modeling of photometric data. To model the AGN components, we used the SKIRTOR model \citep{Skirtor_2016}. It parametrizes AGN UV-OPT emission and physical parameters of the dusty torus responsible for IR emission from the nucleus. The half-opening angle of the torus ($\Delta$) is measured between the equatorial plane and the edge of the dusty surface of the torus. The type I quasars in the SKIRTOR model usually have $\Delta$ values of about 30$^{o}$, and for type II quasars, the value of $\Delta$ is about 70$^{o}$ \citep{Yang2020Xcigale}. CIGALE uses a slightly modified version of SKIRTOR to account for the dust in the polar regions. The additional PD component was added in the work \cite{Yang2020Xcigale}, calculated as a homogeneous screen, with possible different extinction curves \citep{Buat2018}. The exact parameter values used for the SED fitting procedure can be found in Table~\ref{tab:CIGALE_params}. The observational upper limits were taken into account to reject nonphysical models and to better constrain the main physical properties, as was shown, for example, in \cite{Junais2023}.  An example SED plot for one of the quasars in our sample can be seen in Fig. \ref{fig:sed_example}.

\begin{table*}[]
\caption{CIGALE modules and input parameters used for all the fits.}
\label{tab:CIGALE_params}

\begin{tabular}{lll}
\hline
\hline
\multicolumn{1}{|c|}{Parameter}                                                               & \multicolumn{1}{c|}{Symbol}                                           & \multicolumn{1}{c|}{Range}                                       \\ \hline \hline
\multicolumn{3}{|c|}{Delayed Star Formation History and Recent Burst}                                                                                                                                                                    \\ \hline
\multicolumn{1}{|l|}{Age of the main population}                                              & \multicolumn{1}{l|}{age$_{\text{main}}$}                                    & \multicolumn{1}{l|}{4000.0, 6000.0, 9000.0, 12000.0 Myr}                  \\ \hline
\multicolumn{1}{|l|}{e-folding timescale of the delayed SFH}                                  & \multicolumn{1}{l|}{$\tau$}                                                & \multicolumn{1}{l|}{3500.0, 7500.0 Myr}                        \\ \hline
\multicolumn{1}{|l|}{Age of the burst}                                                        & \multicolumn{1}{l|}{age$_{\text{burst}}$}                                   & \multicolumn{1}{l|}{100.0, 300.0 Myr}                          \\ \hline
\multicolumn{1}{|l|}{Burst stellar mass fraction}                                             & \multicolumn{1}{l|}{fburst}                                           & \multicolumn{1}{l|}{0.0, 0.01, 0.1, 0.3}                                               \\ \hline
\multicolumn{3}{|c|}{Dust attenuation based on \cite{CF2000}}                                                                                                                                                                                                   \\ \hline

\multicolumn{1}{|l|}{V-band attenuation in the ISM}                                           & \multicolumn{1}{l|}{A$^{ISM}_{V}$} & \multicolumn{1}{l|}{0.01, 0.3, 0.7, 1.5}                              \\ \hline
\multicolumn{1}{|l|}{Power law slope of dust attenuation in the BCs}                          & \multicolumn{1}{l|}{n$_{BC}$}                                              & \multicolumn{1}{l|}{-0.7 }                                                            \\ \hline
\multicolumn{1}{|l|}{A$^{ISM}_{V}$/(A$^{ISM}_{V}$+A$_{V}^{BC}$)} & \multicolumn{1}{l|}{$\mu$}                                                & \multicolumn{1}{l|}{0.6, 0.44, 0.2}                                         \\ \hline
\multicolumn{1}{|l|}{Power law slope of dust attenuation in the ISM}                          & \multicolumn{1}{l|}{n$_{ISM}$}                                             & \multicolumn{1}{l|}{-1.3, -0.7, -0.48, -0.2}                               \\ \hline
\multicolumn{3}{|c|}{Dust emission \cite{Dale2014} model}                                                                                                                                                                                                    \\ \hline
\multicolumn{1}{|l|}{Alpha}                                                                   & \multicolumn{1}{l|}{$\alpha$}                            &    \multicolumn{1}{l|}{2.0, 2.5, 3.25}                                                              \\ \hline
\multicolumn{3}{|c|}{AGN Skirtor \cite{Skirtor_2016} model\footnote{\textbf{For the fit without polar dust the E(B-V)$_{pd}$ was set only to 0.}}}                                                                                                                                                                                                              \\ \hline
\multicolumn{1}{|l|}{AGN fraction}                                                            & \multicolumn{1}{l|}{f$_{agn}$}                                       &       \multicolumn{1}{l|}{0.7, 0.75, 0.85, 0.9, 0.95, 0.99}                                                           \\ \hline
\multicolumn{1}{|l|}{Opening Angle}                                                           & \multicolumn{1}{l|}{$\Delta$}                            &    \multicolumn{1}{l|}{10, 20, 40, 60, 70}                                                              \\ \hline
\multicolumn{1}{|l|}{Inclination}                                                             & \multicolumn{1}{l|}{$i$}                                              &                     \multicolumn{1}{l|}{10, 20, 40, 60, 80}                                             \\ \hline
\multicolumn{1}{|l|}{Color excess of polar dust}                                              & \multicolumn{1}{l|}{E(B - V)$_{pd}$}                                 &         \multicolumn{1}{l|}{0.0, 0.03, 0.1, 0.15, 0.25, 0.3 (0 for the fit without PD)}                                                         \\ \hline
\multicolumn{1}{|l|}{Temperature of the polar dust}                                           & \multicolumn{1}{l|}{T$_{pd}$}                                        &               \multicolumn{1}{l|}{100.0, 500.0, 1500.0}                                                   \\ \hline
\multicolumn{1}{|l|}{Average edge-on optical depth at 9.7 micron}                                                                        & \multicolumn{1}{l|}{t}                                                 &             \multicolumn{1}{l|}{7, 11}                                                     \\ \hline
\end{tabular}%

\end{table*}

\subsection{Estimates of luminosities}
\label{sec:estimates_lum}
In \cite{Ralowski2024}, we introduced a simple method for calculating observed SED luminosities based on multi-photometric datasets for each galaxy (L$^{photo}_Y$, where $Y$ is the either IR or OPT wavelength range). This quantity is computed as the integral of the SED constructed from observed data points over a given wavelength range. 
The purpose of this method was to provide a practical estimate of luminosity using commonly available filter sets covering the near UV to the MIR. This wavelength range is typically covered by SDSS or other optical telescopes, UKIDSS or 2MASS, and WISE. However, this set may be insufficient for full SED modeling in the absence of UV coverage and reliable MIR and FIR data points. In this work, we perform SED modeling using an extended photometric set that includes GALEX and Herschel data. Our goal is to further calibrate the L$_{photo}$ estimates by comparing them with luminosities derived using the CIGALE code to obtain more reliable results based on the simpler photometric method. The $L^{\text{photo}}_Y$ values were computed via integration in two wavelength ranges: from 0.11 $\mu m$ to 1 $\mu$m the IR luminosity is defined (hereafter L$^{photo}_{IR}$), and between 1 $\mu$m and 7 $\mu$m is the optical luminosity (hereafter L$^{photo}_{OPT}$). We assumed that for AGNs, the dominant contribution to $L_{\text{IR}}$ originates from emission by the dusty torus, while the main component of $L_{\text{OPT}}$ is produced by the accretion disk. 

For the second method, we used best-fit results obtained from SED fitting with CIGALE. The AGN luminosities were estimated from the SKIRTOR model. We extracted the disk luminosity (L$^{disk}_{IR}$) and the torus luminosity estimate (L$^{torus}_{IR}$). All CIGALE luminosities were derived from the best-fit synthetic spectrum and integrated over the same wavelength ranges as in the photometric method. To evaluate contaminating contributions, we included additional components in the $L^{\text{disk}}_{\text{OPT}}$ and $L^{\text{torus}}_{\text{IR}}$ luminosities, specifically stellar emission from the host galaxy, cold interstellar dust, and dust located in the polar regions of the AGN (PD). This resulted in the following luminosities definitions:

\begin{itemize}

    \item  $L_{OPT}^{\text{disk}}$ is based on the AGN disk component, in the wavelength range of 1 $\mu m$ to 7 $\mu$m.
    \item $L_{IR}^{\text{torus}}$ is calculated from the estimated luminosity of the dusty torus, in the wavelength range of 0.11 $\mu m$ to 1 $\mu$m.
    \item $L_{OPT}^{\text{disk+stellar}}$ is the sum of the disk and stellar components, in the wavelength range of 1 $\mu m$ to 7 $\mu$m.
    \item $L_{IR}^{\text{torus+PD}}$ is the sum of the dusty torus and polar dust components, in the wavelength range of 0.11 $\mu m$ to 1 $\mu$m.
    \item $L_{IR}^{\text{torus+PD+disk}}$ is the sum of the dusty torus, polar dust, and accretion disk , in the wavelength range of 0.11 $\mu m$ to 1 $\mu$m.
\end{itemize}

The two independent luminosity estimates enable comparison and calibration of our photometric method. This comparison is shown in Fig.~\ref{fig:Luminosities_vs_contaminations} and is discussed in detail in Sect.~\ref{sec:Lum_comp}. It should be noted that the uncertainties provided by CIGALE for the luminosity estimates are likely overestimated. The CIGALE code computes uncertainties only for bolometric luminosities of each model component. As a result, the reported errors correspond to the full wavelength range of the model, whereas our integration of L$_{IR}$ and L$_{OPT}$ in IR and OPT luminosity ranges are consecutively: 0.11 $\mu m$-1$\mu m$ for IR, and 1 $\mu m$ - 7$\mu m$. This was done for L$_{disk}$ and L$_{torus}$ but also for other components (stellar, cold dust, PD).

\section{Results}
\label{sec:results}

The primary objectives of this analysis is to identify the main sources of contamination that affect the luminosities L$_{\text{IR}}$ and L$_{\text{OPT}}$. Following that to calibrate the photometric luminosity estimation method so that it more accurately reflects the dusty torus and accretion disk luminosities of AGNs as determined through CIGALE fitting.

\subsection{Comparison of luminosity estimations}
\label{sec:Lum_comp}
\begin{figure*}[!htp]

  \includegraphics[clip,width=1\textwidth]{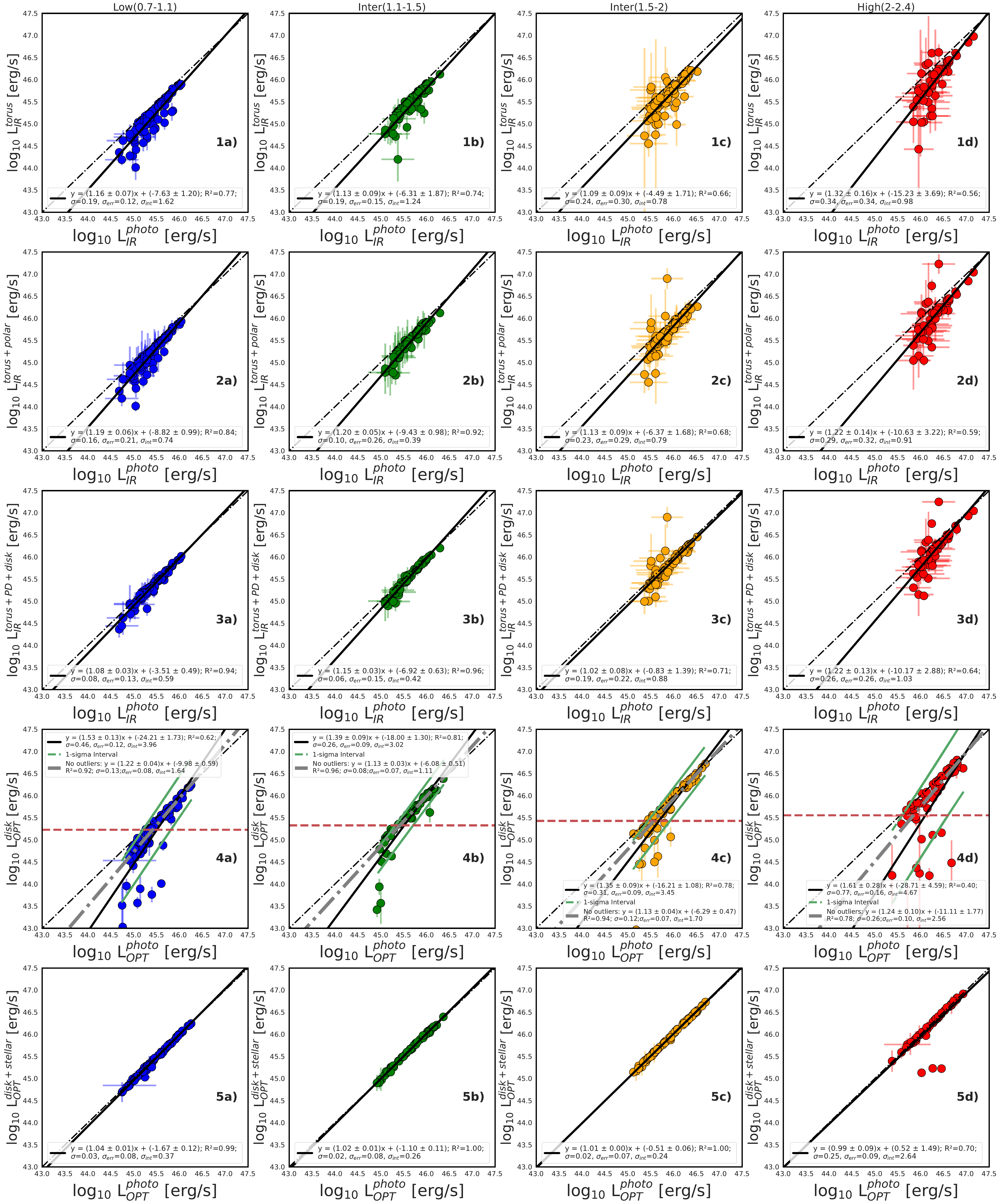}
  
\caption[width=0.5\columnwidth]{Comparison of luminosities estimated from 1) our photometric method (presented on X-axis) and 2) CIGALE estimates without and with contaminations for 4 redshift bins. The dashed line represents the 1:1 relation, while the solid black line corresponds to the best fitted OLS regression. In the fourth row, outliers were identified as objects lying below the $1\sigma$ threshold (indicated by green lines) and are marked with star symbols. These outliers were excluded, and a second OLS regression was fitted to the remaining data points (gray dash-dot line). Additionally, in the fourth row, the red dashed line indicates the luminosity cut described in Section~\ref{sec:Lum_Cut}. The $\sigma$ is the spread of the relation.
}
\label{fig:Luminosities_vs_contaminations}
\end{figure*}

The integrated luminosities were computed using two distinct approaches: 1) from the photometric SED, yielding L$^{photo}_{IR}$ and L$^{photo}_{OPT}$, and 2) based on the CIGALE estimations, L$^{torus}_{IR}$ and L$^{disk}_{OPT}$. The former estimations were subsequently combined with the dominant contaminating components. 
In Fig. \ref{fig:Luminosities_vs_contaminations} the comparison between both estimates is shown. The fitter regression is the ordinary least squares regression. The spread ($\sigma$) was calculated as the standard deviation of residuals from the regression fit. Further, the mean uncertainty of observations ($\sigma_{err}$) was calculated as $\sigma_{err} = mean(\sqrt{x_{err}^{2}+y_{err}^2})$, where $x_{err}, y_{err}$ are the observation error of X and Y variables, respectively. To test what the contribution of the spread from the observational uncertainties is within the total spread of the relation, the intrinsic scatter ($\sigma_{int}$) was calculated as ratio of $\sigma/\sigma_{err}$. The subsequent rows in Fig. \ref{fig:Luminosities_vs_contaminations} show different sources of contamination that affect the CIGALE-based estimations.
A tight correlation between the two luminosity estimates, in the IR and OPT ranges, can clearly be observed. As expected, the photometric estimates exhibit systematically higher values and effectively serve as upper limits to the CIGALE-based estimates. The observed offset from the 1:1 relation, along with the scatter in the L$^{torus}_{IR}$ versus L$^{photo}_{IR}$ relation (Fig. \ref{fig:Luminosities_vs_contaminations}, first row) indicates the presence of additional IR contributions that cannot be attributed solely to polar dust (Fig. \ref{fig:Luminosities_vs_contaminations}, second row). Our analysis shows that the accretion disk emission extends beyond the OPT range and contributes significantly to the IR regime as well. As shown in Table~\ref{tab:comp_high}, the disk contribution to L$^{photo}_{IR}$ is significant. After inclusion of L$_{IR}^{torus+PD+disk}$, the relation is not only closer to 1:1, but it also shows a narrower spread (Low-$z$ 0.16 - 0.08, Inter1-$z$ 0.10 - 0.06, Inter2-$z$ 0.23-0.19, High-$z$ 0.29 - 0.26; see the third row in Fig. \ref{fig:Luminosities_vs_contaminations}). The remaining discrepancy between the data points and the 1:1 relation is attributed primarily to stellar emission and cold dust. In general, for redshifts below 1.5, the intrinsic scatter drops (from $\sim$1.6 to 0.7 for Low-$z$, $\sim$1.2 to 0.4 for Inter1-$z$), while for higher redshifts, the $\sigma_{int}$ slightly increases.

In the OPT range, the correlation between the modeled disk emission and the photometric estimates follows a well-defined main trend, as illustrated in Fig.~\ref{fig:Luminosities_vs_contaminations}, fourth row. Additionally, the L$^{disk}_{OPT}$ versus L$^{photo}_{OPT}$ relation also exhibits a substantial number of outliers across all redshift bins. To investigate how these outliers differ from the main quasar population, we conducted two dedicated tests.   
The first test involved fitting two least-squares regression lines: one to the full sample of quasars (solid black line, shown in Fig.~\ref{fig:Luminosities_vs_contaminations}, fourth row) and a second excluding outliers.  
Outliers were defined as objects lying beyond a 1$\sigma$ threshold from the first regression line. The second regression, fit to the remaining sources, is represented by the dash-dot gray line and follows the main trend more closely, with a slope approaching unity (Low-$z$ from 1.53$\pm$0.13 to 1.22 $\pm$ 0.04; Inter1-$z$ from 1.39$\pm$0.09 to 1.13$\pm$0.03; Inter2-$z$ from 1.35$\pm$0.09 to 1.13$\pm$0.04; High-$z$ from 1.61$\pm$0.28 to 1.24$\pm$0.10). When the stellar component is included (Fig.~\ref{fig:Luminosities_vs_contaminations}, fifth row), the relation between the photometric and CIGALE luminosities becomes consistent with a 1:1 relation, within the uncertainties, particularly for the previously identified outliers. A subset of these outliers can thus be attributed to a significant stellar contribution, which is absent in L$^{disk}_{OPT}$ but is present in the next row of panels with L$^{disk+stellar}_{OPT}$. Most of the remaining sources exhibit high f$_{agn}$values, indicating that stellar emission plays a less significant role in their total luminosity.
The ratio of each contamination can be seen in Tables \ref{tab:comp_high} and \ref{tab:comp_low}. The outliers were responsible for the majority of the $\sigma_{int}$, as it drops massively for each redshift range.

It should noted that the low-luminosity outliers with a higher AGN fraction are influenced by the addition of polar dust in L$_{IR}$ (see Inter-$z1$), visible in Fig. \ref{fig:Comparison_OPT-IR_appendix}. For outliers with a lower AGN fraction, the stellar component plays a significant role in estimating the L$_{OPT}$.

\subsection{Luminosity cut}
\label{sec:Lum_Cut}

The quasars shown in the fourth row of Fig.~\ref{fig:Luminosities_vs_contaminations} exhibit increased scatter at the low-luminosity end of L$^{disk}_{OPT}$. To have a universal selection of objects, we applied a luminosity cut at L$^{disk}_{OPT}\sim10^{45}$ across all redshifts, similar to the threshold adopted by \cite{Trefoloni2024}. The spread $\sigma$ for the objects below L$^{disk}_{OPT}=10^{45}$ is almost ten times wider compared to the quasars above the threshold (low-$z$: 0.63 - 0.07, inter1-$z$ 0.36 - 0.08, inter2-$z$ 0.51 - 0.06, high-$z$ 1.06 - 0.11 for objects below and above the L$^{disk}_{OPT}=10^{45}$ threshold).
This separation divides the sample into two groups of objects: 1) lower-luminosity sources with greater scatter and stronger contamination and 2) brighter objects with a more robust trend. Our goal was to examine whether different types of contamination affect these two regimes in distinct ways
As shown in Fig.~\ref{fig:Luminosities_vs_contaminations}, a fixed threshold of L$^{disk}_{OPT}=10^{45} [erg/s]$ does not perfectly follow the "break point" of the OPT-IR relation for all redshift due to luminosity evolution. Therefore, we propose a redshift-dependent cut, defined as  L$^{disk}_{OPT}=10^{45}+0.25\times|z|$, where $|z|$ is the mean redshift in each bin. The selection of this particular threshold was mainly empirical, as proper selection should be based on fluxes. This issue is discussed in \ref{sec:Discussion} and in Fig. \ref{fig:Lum_Cut_Redshift}.
This adjusted threshold serves two main purposes: 1) to identify the luminosity regime in which both methods, photometric and CIGALE estimated, yield consistent OPT–IR relations and 2) to calibrate the photometric method in a region of reduced scatter. 
Although a similar effect can be observed in the IR part, we did not impose an additional cut for it.

\subsection{Comparison of OPT-IR relation, through different luminosity estimations}
\label{sec:OPT_IR_estimations}

\begin{figure*}[!htp]

  \includegraphics[clip,width=1\textwidth]{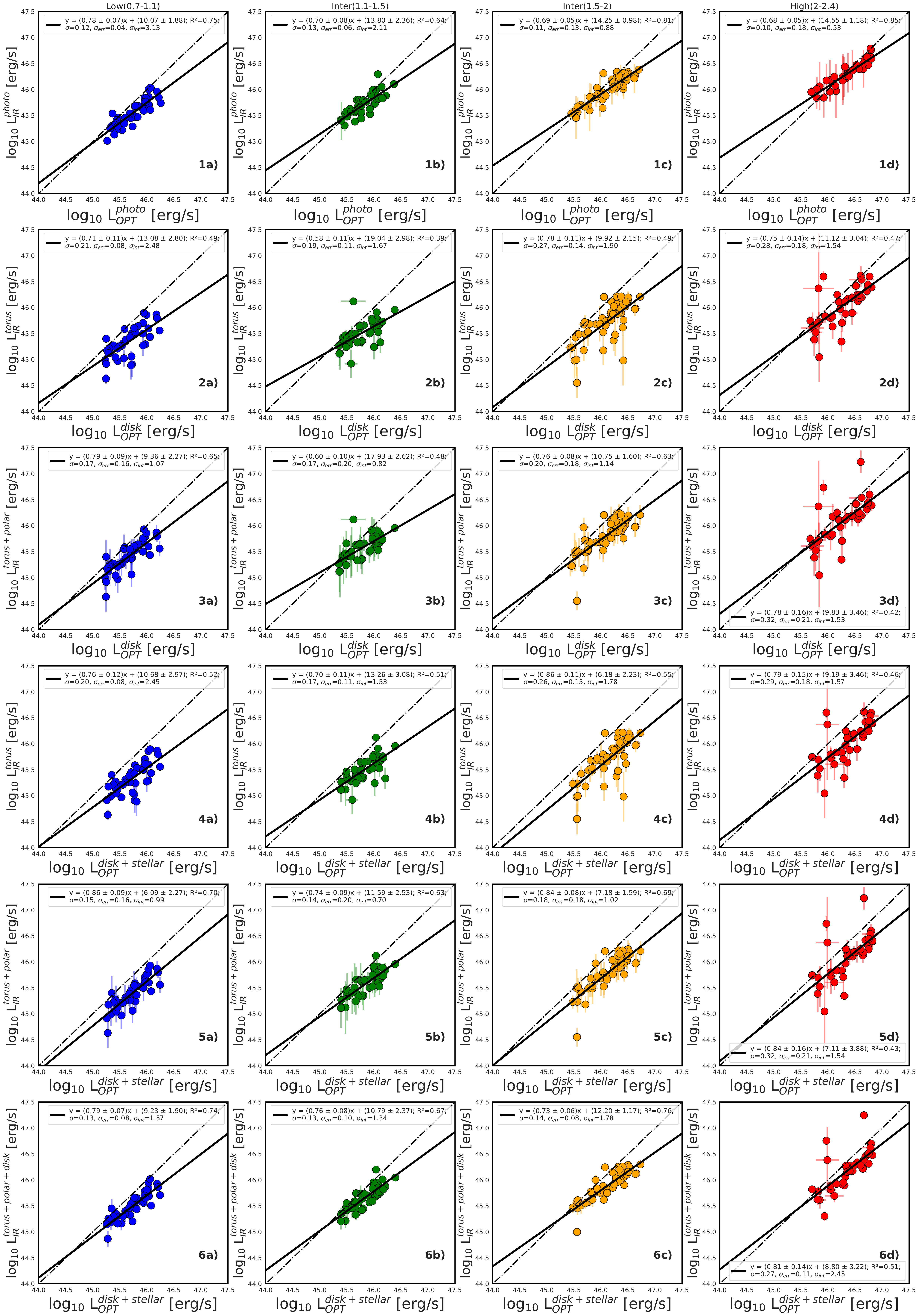}
  
\caption[width=0.5\columnwidth]{The OPT-IR relation for all redshift bins through different estimates of luminosities. The black solid lines are the OLS regressions. In the first row the $L_{IR}$ and $L_{OPT}$ are estimated with the photometric method. The legend is explained under previous plot.}
\label{fig:Comparison_OPT-IR}
\end{figure*}

The OPT–IR relation between L$_{photo}^{IR}$ and L$_{photo}^{OPT}$ exhibits the strongest correlation ($R^2\simeq0.8$) and the narrowest spread ($\sigma\simeq0.11$). Uncertainties for the photometric method were estimated using Monte Carlo Markov chain simulations. These uncertainties increase with redshift, reflecting the larger observational errors associated with higher redshift bins. The errors for the SED fitting estimations are overestimated, as they are computed over the entire model wavelength range, rather than being limited to the relevant IR or OPT intervals (as noted in Sect.~\ref{sec:SED}). The second row of Figure~\ref{fig:Comparison_OPT-IR} shows the relation between L$^{disk}_{OPT}$ and L$^{torus}_{IR}$. A low-luminosity tail is clearly visible, characterized by increased scatter in L$^{disk}_{OPT}$, particularly among low-$z$, low-luminosity objects. The third row shows that the addition of polar dust luminosity in L$^{\text{torus + polar\ dust}}_{\text{IR}}$ does not significantly alter the distribution, but it does introduce large uncertainties for most objects. These uncertainties are again calculated over the full wavelength range of each model since CIGALE does not provide error estimates for individual wavelength intervals (e.g., IR or OPT) separately. Consequently, the resulting uncertainties are likely overestimated. The uncertainties for the sum of luminosities were computed via standard error propagation methods.

In contrast, the inclusion of the stellar luminosity component in L$^{disk + stellar}_{OPT}$ substantially increases the total L$_{OPT}$, exerting a significant influence on the group of outliers previously observed in the panel showing only L$^{disk}_{OPT}$. Uncertainties for high-redshift objects remain the largest among the sample. The luminosity distributions that account for the contamination sources L$^{disk + stellar}_{OPT}$, L$^{torus + PD + disk}_{IR}$ seem to be the most reminiscent of the photometric method, which is confirmed by the slopes of the fit regressions.

Surprisingly, the distribution of the AGN fraction is mixed even among the most luminous objects. No clear correlation was observed between the AGN fraction and the total luminosity. Notably, some of the highest-luminosity sources exhibit low AGN fractions, as illustrated in Fig.~\ref{fig:Comparison_OPT-IR_appendix}. This can be explained by the imperfections in the SED fitting, as after checking the PDF and fit half-opening angles, we found that most of these objects were classified as type II quasars. However, for the majority of these cases, the type I quasar solutions yielded comparable $\chi^2$ values and were accompanied by lower stellar luminosity components, higher AGN fractions, and smaller opening angles. We tested whether the low AGN fraction might be due to a higher starburst fraction (f$_{*}$). For the majority of objects, the average value of f$_{*}$ is approximately f$_{*} \sim 0.15$, which is similar to the main group of objects.

The OPT-IR relation for photometric luminosities in the first row of Fig. \ref{fig:Comparison_OPT-IR} shows significant similarities with those in the fifth row regarding all major contaminations in terms of both regression equation and the spread of the relation. The only exception is the high-$z$ sample. This signifies that within the OPT-IR relation, there are indeed contaminations influencing the distribution. We address this issue in the next section.

\section{Contamination analysis}
In Sect.~\ref{sec:Lum_Cut} we demonstrated that the photometric method is prone to contamination, especially for the objects with a low L$_{OPT}$. Therefore, to reliably investigate the relationship between the accretion disk and the torus in AGNs, it is essential to account for multiple sources of contamination, including polar dust within the AGN, stellar emission, cold dust from the host galaxy,and disk emission extending into the IR. In this section we analyze each of these major contamination sources in detail, taking into consideration their potential redshift-dependent effects.
Tables~\ref{tab:comp_high} and \ref{tab:comp_low} present a comparison between CIGALE-derived luminosity components (with additional contaminations) and the corresponding L$^{photo}_{IR}$ and L$^{photo}_{OPT}$ values. It is worth noting that L$^{torus}_{IR}$ and L$_{OPT}^{disk}$ respectively constitute the dominant contributions to L$^{photo}_{IR}$ and L$^{photo}_{OPT}$, with their fractional contributions increasing at higher redshifts (Table~\ref{tab:comp_high} rows (1) and (5)).

\subsection{Polar dust}
One of the key sources of contamination analyzed in this work is dust located in the polar regions of the AGN, commonly referred to as PD. To investigate its impact, we performed SED fitting under two configurations: 1) with polar dust with parameters described in Table \ref{tab:CIGALE_params} and 2) without polar dust, i.e., parameter E(B-V)$_{PD}$ was set to 0 (see also Table \ref{tab:CIGALE_params}). The main results of this comparison are discussed in Sect.~\ref{sec:no_polar}.

\subsubsection{Fit with PD}
Polar dust was modeled within the SED fitting framework using CIGALE, specifically through the SKIRTOR module. In this implementation, the SKIRTOR module was slightly modified by incorporating a polar dust component, represented as a blackbody emitter (for further details, see \cite{Yang2020Xcigale}). Polar dust has a relatively minor influence on L$_{\text{IR}}$. For the brighter objects, 
the polar dust luminosity is up to a few percent for all redshift ranges. For objects below the luminosity threshold, this contamination is higher (8\% for low-$z$, 3\% for inter1-$z$, 4\% for inter2-$z$, and 9\% for high-$z$).

Additionally, we examined the influence of polar dust by analyzing the ratio L$_{\text{IR}}^{\text{torus}}$/L$^{\text{PD}}_{\text{IR}}$. This metric quantifies the extent to which the torus luminosity dominates over the polar dust contamination. As shown in Fig.~\ref{fig:all_ratio}, the torus is typically found to be four to six times more luminous (on average 79\% across all redshift samples with L$_{\text{IR}}^{\text{torus}}$/L$^{\text{PD}}_{\text{IR}}$ greater than four and 40\% when above five) than the polar dust component (indicated in brown and pink colors). 
Among the objects below the luminosity threshold, there are more objects with a lower L$_{\text{torus}}$/L$_{\text{polar dust}}$ ratio (only 7\% objects above five). The only sample different from this trend is the inter2-$z$ at\ Fig. \ref{fig:all_ratio}, where the majority of objects have a very high ratio. 
Objects above and below the luminosity threshold appear to exhibit similar values of the L$_{\text{IR}}^{\text{torus}}$/L$^{\text{PD}}_{\text{IR}}$ ratio. Although the most luminous sources with L$^{photo}_{OPT}>10^{45}$ [erg/s] tend to have the highest ratio, the torus remains the dominant IR component by a substantial margin. This trend seems to be consistent across redshift bins.

When comparing the reddening E(B-V)$_{PD}$ caused by the presence of polar dust on the line of sight, for the objects  below L$^{disk}_{OPT} \sim 10^{45}$ across the full redshift range, the polar dust has a higher E(B-V)$_{polar\ dust}$ value of $~0.18$, which is over two times greater than objects above the luminosity threshold. The high-$z$ sample below the threshold has the lowest E(B-V)$_{polar\ dust}$ of around 0.14. This is particularly evident in Fig.~\ref{fig:All_E(B-V)} (lower panel), where objects below the threshold show increased dust attenuation. The estimated temperature of the polar dust in the low-$z$ bin is, on average, the lowest among all redshift groups, with a median value of 170,K $\pm$ 70,K, and it strongly affects the FIR portion of the spectrum. Inter1-$z$ objects have a high E(B-V)$_{PD}$ (fig. \ref{fig:All_E(B-V)}). These results confirm that the luminosity cut effectively excludes sources with significant polar dust attenuation, not only at low redshift but also in higher redshift bins. A similar trend is observed for inter2-$z$ objects, where those with the highest E(B-V)$_{polar\ dust}$ have L$^{disk}_{OPT}<10^{45} [erg/s]$ (see Fig. \ref{fig:All_E(B-V)}). It should be noted that the E(B-V)$_{polar\ dust}$ can be influenced not only by the amount of the polar dust but also by the inclination, which can significantly change its values.

\subsubsection{Fit without polar dust}
\label{sec:no_polar}
 In the previous section, we described the characteristics of the SED fitting we performed with the inclusion of polar dust. As shown in \cite{Toba_2021}, the inclusion or exclusion of polar dust can significantly affect the derived model parameters. Motivated by this analysis, we conducted a second SED fitting run with polar dust excluded by setting E(B-V)$_{pd}$=0 (see Table~\ref{tab:CIGALE_params}). The sample size after applying a reduced $\chi^2$ cut was smaller: 50, 40, 50, and 38 sources respectively for the low-$z$, inter1-$z$, inter2-$z$, and high-$z$ samples. In this configuration, the torus luminosity increased by up to 8\% of the IR emission, while the disk contribution to the IR remained similar to the previous fit, at approximately 16\%. (For the results of the fit without polar dust, see Tables~\ref{tab:No_polar_above} and \ref{tab:No_polar_below}.)
 A visual inspection of the SEDs indicated that for objects previously fit with hot polar dust, the absence of this component leads to torus models with systematically higher temperatures. 
From the comparison between Figs.~\ref{fig:Luminosities_vs_contaminations} and \ref{fig:No_polar_Luminosities_vs_contaminations} in the $L^{torus}_{IR}$ versus $L_{IR}^{photo}$ (first row), one can see that 1) the torus luminosity is slightly higher in the model without polar dust, and 2) the overall scatter remains comparable across all redshift bins (on average $\sigma=0.03$ compared to $\sigma=0.04$ for the fit with polar dust). Additionally, the number of outliers based on the regression fit changed (13-4-11-4, for low-inter1-inter2-high datasets, compared to 8-8-9-10).

\begin{figure*}[!htp]

  \includegraphics[clip,width=0.8\textwidth]{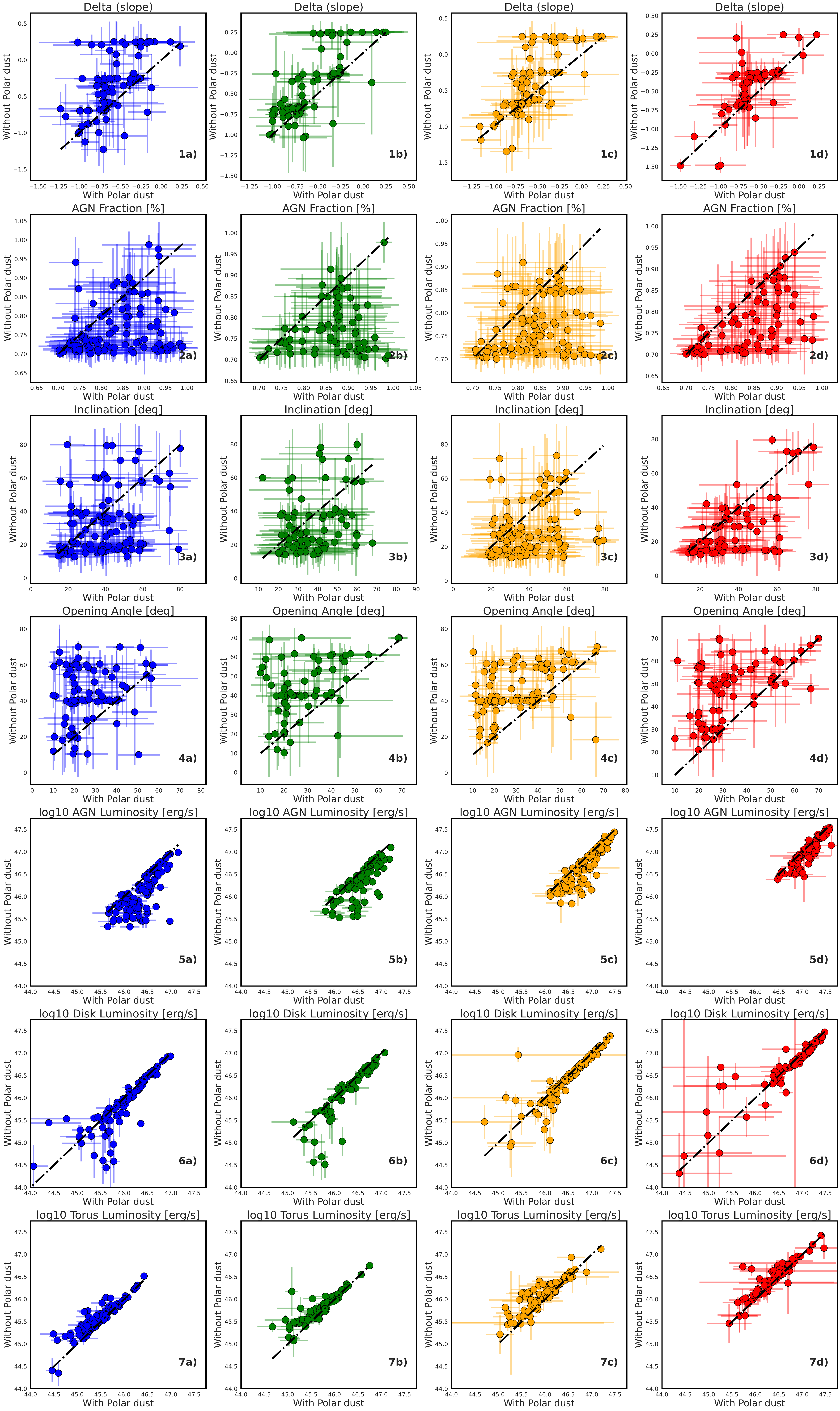}
  
\caption[width=0.5\columnwidth]{Comparison of SKIRTOR model parameters derived from two independent SED fitting runs: 1) With polar dust, and 2) without polar dust ($E(B-V)_{PD}=0$). The X-axis shows values from the fit including polar dust, while the Y-axis corresponds to the fit without polar dust. The plotted parameter is indicated in the panel title. The black dash-dotted line represents the 1:1 relation.
}
\label{fig:with_vs_out}
\end{figure*}

A detailed comparison between the two SED fitting approaches is presented in Fig.~\ref{fig:with_vs_out}. The best-fit SKIRTOR parameters differ significantly between the fits with and without polar dust. 
 The fit without PD is characterized by the following parameter differences: 1) a steeper slope of the accretion disk emission, $\delta$; 2) a significantly lower AGN fraction, $f_{agn}$, with an average value of $\sim$70\% (see Fig.~\ref{fig:with_vs_out}, second row); 3) a lower average inclination angle ($i$), approximately $\sim 30$ [deg] compared to $\sim 40$ [deg] in the fit with PD; 4) a higher opening angle, with a median of 44 [deg] compared to 30 [deg] when PD is included; and 5) the total accretion disk luminosity (fifth row in Fig.~\ref{fig:with_vs_out}) has higher errors for low luminous objects. The total AGN luminosity is slightly higher in the fit that includes polar dust, whereas the torus luminosity increases in the fit without polar dust. 

\begin{figure*}[!htp]

  \includegraphics[clip,width=1\textwidth]{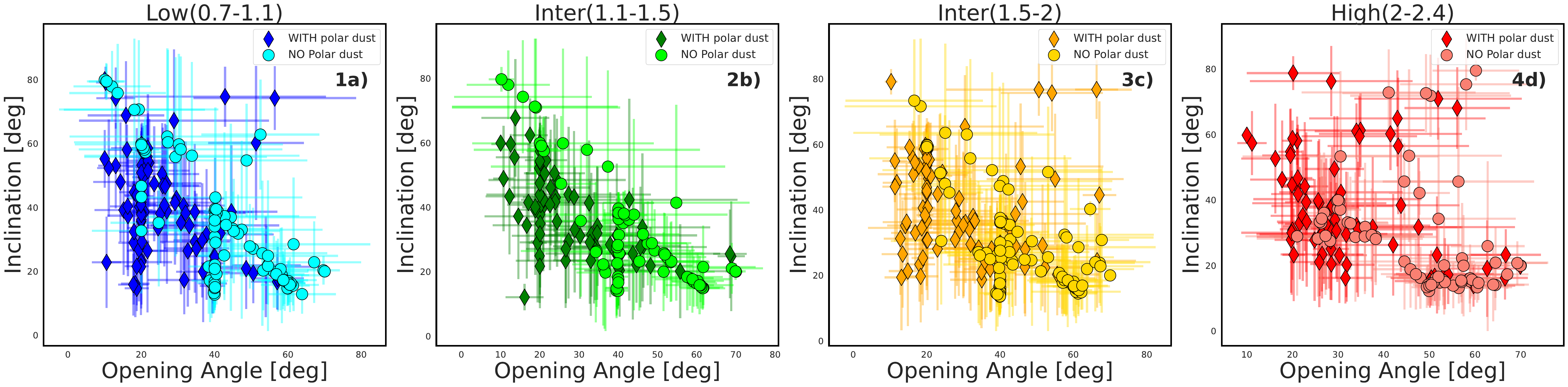}

\caption[width=0.8\textwidth]{Relation between OA on X axis and inclination on Y axis for fit with polar dust (diamonds) and without (circles). %
With grey arrows the shift for each object for both fits is marked.

}
\label{fig:OA_i_PD}
\end{figure*}

\begin{figure*}[!htp]

  \includegraphics[clip,width=1\textwidth]{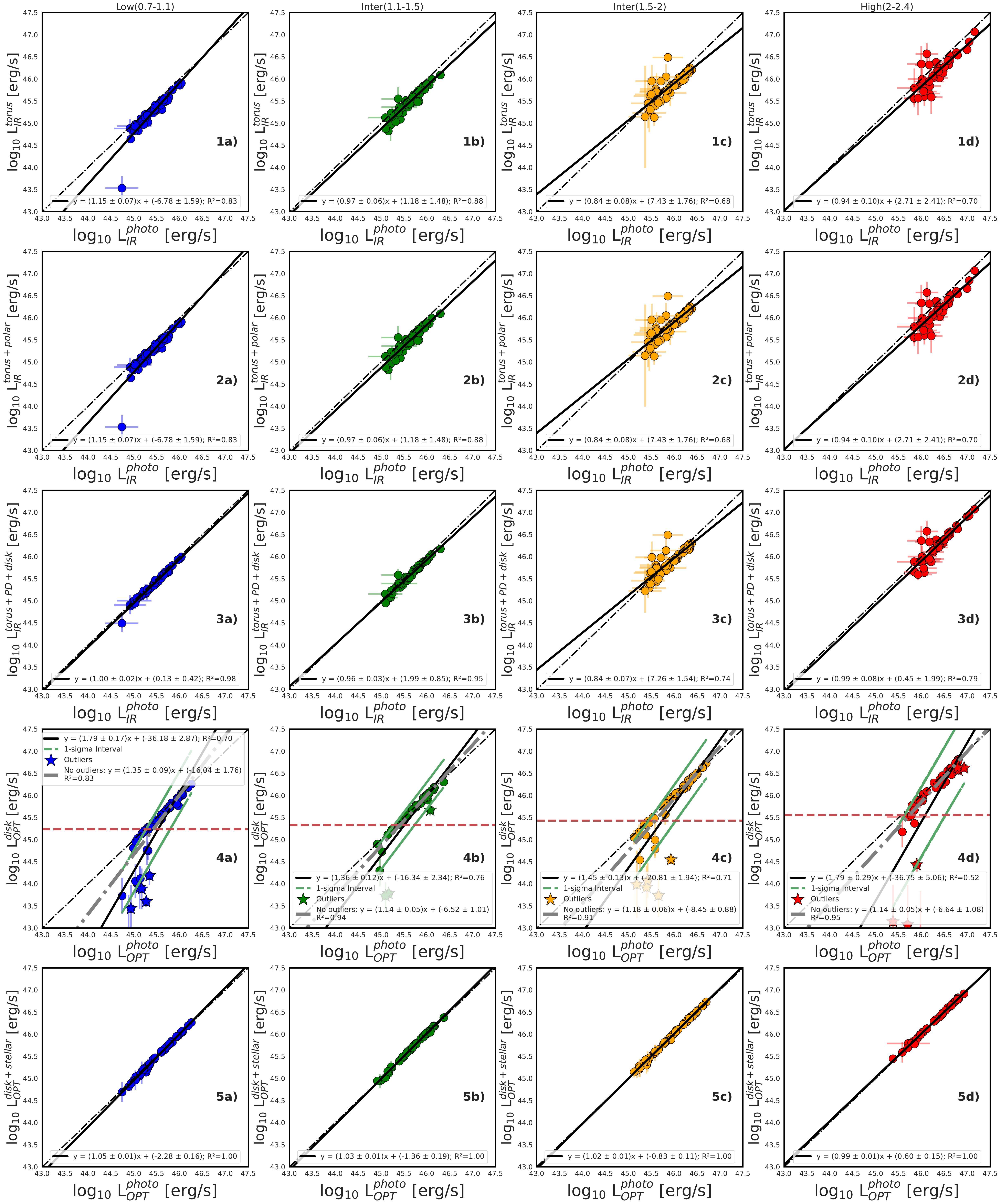}
  
\caption[width=0.5\columnwidth]{No polar-dust: Comparison of luminosities estimated from 1) our photometric method (presented on X-axis) and 2) CIGALE estimates without and with contaminations for 4 redshift bins. The dashed lines is the 1:1 relation, while the solid, black line is the best fitted regression. In the 4th row the outliers were also accounted for, defined as objects below 1$\sigma$ error (green lines), marked with stars. Then the outliers were extracted and another regression was fitted (grey dashed line).}
\label{fig:No_polar_Luminosities_vs_contaminations}
\end{figure*}

Overall, considering both this result and the opening angle (panel 4, Fig.\ref{fig:with_vs_out}), the majority of objects fit without polar dust can be interpreted as type II objects (low AGN fraction, high opening angle). In general, the inclusion of polar dust affects both the opening angle and inclination angle ($i$). For most objects, the fit opening angle values decrease when polar dust is included, as illustrated in Fig.~\ref{fig:OA_i_PD}. For redshifts in the range $0.7 < z < 2.0$, a clear degeneracy is observed in the fit — manifested as a triangular distribution at an opening angle of around 40 [deg] and a diagonal trend. For high-$z$ objects, the fit opening angle values tend to increase (typically reaching $\sim$50--60 [deg]), while the inclination angles decrease.

\begin{table*}[htbp]
\centering
\caption{Median values of ratios of different estimations of luminosities for objects with L$_{disk}^{OPT}>10^{45}$ [erg/s]. The errors are the MAD errors. The particular luminosities estimates are described in section \ref{sec:OPT_IR_estimations}.  
}

\begin{tabular}{|c|c|c|c|c|}
\hline
Column & Low & Intermediate 1  & Intermediate 2  & High \\
\hline
(1)L$^{torus}_{IR}$/L$^{photo}_{IR}$ & 0.73 $\pm$ 0.14 & 0.72 $\pm$ 0.12 & 0.70 $\pm$ 0.16 & 0.71 $\pm$ 0.14 \\
(2) L$_{IR}^{\text{torus+polar}}$/L$_{IR}^{\text{photo}}$ & 0.76 $\pm$ 0.07 & 0.76 $\pm$ 0.06 & 0.76 $\pm$ 0.09 & 0.75 $\pm$ 0.10 \\
(3) L$_{IR}^{\text{torus+polar+cold}}$/L$_{IR}^\text{{photo}}$ & 0.78 $\pm$ 0.06 & 0.79 $\pm$ 0.05 & 0.79 $\pm$ 0.08 & 0.77 $\pm$ 0.09 \\
(4) L$_{IR}^{\text{torus+polar+cold+stellar}}$/L$_{IR}^\text{{photo}}$ & 0.83 $\pm$ 0.06 & 0.83 $\pm$ 0.05 & 0.83 $\pm$ 0.08 & 0.84 $\pm$ 0.08 \\
(5) L$_{IR}^{\text{torus+polar+cold+stellar+disk}}$/L$_{IR}^\text{{photo}}$ & 1.00 $\pm$ 0.00 & 1.00 $\pm$ 0.00 & 1.00 $\pm$ 0.00 & 1.00 $\pm$ 0.00 \\
(6) L$_{IR}^{torus+polar}$/L$_{IR}^{torus}$ & 1.00 $\pm$ 0.44 & 1.00 $\pm$ 0.33 & 1.00 $\pm$ 0.76 & 1.00 $\pm$ 0.37 \\
(7) L$_{OPT}^{disk}$/L$^{photo}_{OPT}$ & 0.85 $\pm$ 0.11 & 0.94 $\pm$ 0.08 & 0.94 $\pm$ 0.09 & 0.85 $\pm$ 0.13 \\
(8) L$_{OPT}^{disk+stellar\ attenuated}$/L$^{photo}_{OPT}$ & 0.99 $\pm$ 0.01 & 1.00 $\pm$ 0.01 & 1.00 $\pm$ 0.01 & 1.00 $\pm$ 0.01 \\
(9) L$_{OPT}^{disk+stellar\ attenuated}$/L$_{IR}^{disk}$ & 1.17 $\pm$ 0.17 & 1.06 $\pm$ 0.14 & 1.07 $\pm$ 0.12 & 1.18 $\pm$ 0.25 \\

\hline
\end{tabular}
\label{tab:comp_high}
\end{table*}

\begin{table*}[htbp]
\centering
\caption{Median values of ratios of different estimations of luminosities for objects with L$_{disk}^{OPT}<10^{45}$ [erg/s]. The errors are the MAD errors. The particular luminosities estimates are described in section \ref{sec:OPT_IR_estimations}}

\begin{tabular}{|c|c|c|c|c|}
\hline
Column & Low (45) & Intermediate 1 (30)  & Intermediate 2 (22) & High (17) \\
\hline
 
(1)L$^{torus}_{IR}$/L$^{photo}_{IR}$ & 0.55 $\pm$ 0.17 & 0.57 $\pm$ 0.14 & 0.64 $\pm$ 0.16 & 0.72 $\pm$ 0.26 \\
(2) L$_{IR}^{\text{torus+polar}}$/L$_{IR}^{\text{photo}}$ & 0.63 $\pm$ 0.16 & 0.60 $\pm$ 0.12 & 0.68 $\pm$ 0.15 & 0.81 $\pm$ 0.13 \\
(3) L$_{IR}^{\text{torus+polar+cold}}$/L$_{IR}^\text{{photo}}$ & 0.71 $\pm$ 0.15 & 0.72 $\pm$ 0.10 & 0.75 $\pm$ 0.13 & 0.83 $\pm$ 0.12 \\
(4) L$_{IR}^{\text{torus+polar+cold+stellar}}$/L$_{IR}^\text{{photo}}$ & 0.83 $\pm$ 0.14 & 0.79 $\pm$ 0.11 & 0.85 $\pm$ 0.13 & 0.96 $\pm$ 0.06 \\
(5) L$_{IR}^{\text{torus+polar+cold+stellar+disk}}$/L$_{IR}^\text{{photo}}$ & 1.00 $\pm$ 0.00 & 1.00 $\pm$ 0.00 & 1.00 $\pm$ 0.00 & 1.00 $\pm$ 0.00 \\
(6) L$_{IR}^{torus+polar}$/L$_{IR}^{torus}$ & 1.00 $\pm$ 0.37 & 1.00 $\pm$ 0.88 & 1.00 $\pm$ 0.64 & 1.00 $\pm$ 2.16 \\
(7) L$_{OPT}^{disk}$/L$^{photo}_{OPT}$ & 0.53 $\pm$ 0.25 & 0.76 $\pm$ 0.30 & 0.50 $\pm$ 0.26 & 0.38 $\pm$ 0.28 \\
(8) L$_{OPT}^{disk+stellar\ attenuated}$/L$^{photo}_{OPT}$ & 0.97 $\pm$ 0.02 & 0.98 $\pm$ 0.03 & 0.98 $\pm$ 0.02 & 0.96 $\pm$ 0.02 \\
(9) L$_{OPT}^{disk+stellar\ attenuated}$/L$_{IR}^{disk}$ & 1.82 $\pm$ 62.94 & 1.31 $\pm$ 5.91 & 1.95 $\pm$ 13.29 & 2.53 $\pm$ 713.33 \\

\hline
\end{tabular}
\label{tab:comp_low}
\end{table*}

\subsection{Cold dust}
Cold dust from the ISM has a minor influence on  L$^{\text{photo}}_{\text{IR}}$, as the majority of its emission lies beyond the upper wavelength limit of 7~$\mu$m used in our integration. This FIR component contributes approximately 3\% to L$^{photo}_{IR}$ for objects above the luminosity threshold, across all redshift bins. For sources below the threshold, the impact is more pronounced: $\sim$ 8\% for low-$z$, 12\% for inter1-$z$, 7\% for inter2-$z$, and 2\% for high-$z$. In the fit without polar dust, the contribution of cold dust is reduced to around 2\% across all redshifts. This is due to the dusty torus effectively absorbing the role of both the polar dust and cold dust components in the IR range.

\subsection{Stellar emission}
Stellar emission constitutes the primary source of contamination in L$^{\text{disk}}_{\text{OPT}}$ across all redshift bins. This effect is clearly visible in Fig.~\ref{fig:Luminosities_vs_contaminations}, fourth row. All outliers (marked with stars) exhibiting a low L$^{\text{disk}}_{\text{OPT}}$ show a substantial stellar component. When stellar emission is added and the relation between L$^{\text{disk+stellar}}_{\text{OPT}}$ and L$^{\text{photo}}_{\text{OPT}}$ is plotted, a tight correlation emerges, and the previously observed outliers disappear. The slope of the fit regression approaches unity for all redshift intervals. A lower stellar contribution is typically observed for sources with L$^{\text{disk}}_{\text{OPT}}>10^{45}$ [erg/s]. For these brighter objects, the average stellar contamination is approximately 5\% (5\% Low-$z$, 4\% Inter1-$z$, 4\% Inter2-$z$, and 7\% High-$z$). In contrast, for dimmer sources, the stellar contamination increases to around 10\%, with even higher values for $z > 1.5$: 12\% Low-$z$, 7\% Inter1-$z$, 10\% Inter2-$z$, and 13\% High-$z$. This trend is likely due to the reduced quality of UV photometric data at higher redshifts, which is critical for accurately separating AGN and stellar components in the UV/optical range. Overall, stellar emission remains the dominant source of contamination in $L_{\text{OPT}}$ across the full sample. In the fit without polar dust, stellar contamination levels are similar for sources above the luminosity threshold: 6\% for low-$z$, 5\% for inter1-$z$, 6\% for inter2-$z$, and 4\% for high-$z$. However, for sources below the threshold, contamination is higher: 9\% for low-$z$, 12\% for inter1-$z$, 6\% for Inter2-$z$, and 9\% for high-$z$.

\subsection{Accretion disk emission in IR luminosity}
The accretion disk emission modeled in SKIRTOR extends into the IR regime. This component of the disk continuum is visible in Fig.~\ref{fig:sed_example} as an dashed orange line for the example SED. The calculated photometric luminosity is directly influenced by the definition of the wavelength ranges used for OPT and IR luminosities. Since the accretion disk emits across a broad spectral range, including the near-IR, its contribution to $L^{\text{photo}}_{\text{IR}}$ can be significant. Surprisingly, for sources above the luminosity threshold, the disk contributes approximately 17\% to L$^{disk}_{IR}$. For dimmer objects, those below the threshold (L$^{disk}_{OPT}<10^{45}$ [erg/s]), the contribution from disk emission to IR decreases with redshift: 17\% low-$z$, 21\% inter1-$z$, 15\% Inter2-$z$, and 4\% high-$z$. 
In the fit without polar dust, L$^{disk}_{IR}$ maintains a similar contribution of $\sim$17\% to L$^{photo}_{IR}$ for brighter objects (see Table~\ref{tab:No_polar_above}). For sources below the luminosity threshold, the disk contamination in IR shows a strong redshift dependence:
21\% for low-$z$ and decreasing to 1\% for high-$z$ (see Table~\ref{tab:No_polar_below}).

\subsection{AGN properties}
To further investigate the differences between sources below and above the L$^{disk}_{OPT} \sim 10^{45}$ [erg/s] threshold and SED fits performed with and without the polar dust component, we analyzed and compared the physical properties of the quasars. Our analysis focused primarily on AGN-related parameters as modeled by the SKIRTOR framework.

The vast majority of sources exhibit torus opening angle, $\Delta$, values between 20 [deg] and 40 [deg], as can be seen in Fig. \ref{fig:Median_oa_er_full}, with an average of 30 [deg]. A systematic shift in the opening angle inclination, $\Delta$ - i, space can be seen in Fig. \ref{fig:OA_i_PD}. Fits that include polar dust tend to yield smaller opening angles and larger inclination angles compared to those without polar dust. 
For fainter sources, the fit parameters are associated with significantly larger uncertainties,
which can be attributed to two main factors: 1) lower flux levels and consequently higher observational errors and 2) an increased level of contamination, as discussed in the previous section. 

We found that high-luminosity objects in high-$z$ have a high opening angle ($\Delta \sim 70$ [deg]) and high $L_{IR}>10^{47} [erg/s]$, suggesting a possible type II AGN and  an ultraluminous IR galaxy. In general, high-$z$ objects above the luminosity threshold have higher $\Delta$ values, as can be seen in Fig. \ref{fig:high_OA}. Additionally, in Fig. \ref{fig:high_delta}, the luminosity relation with the power-law of index $\delta$ is presented. 
The higher values of the $\delta$ slope imply a redder spectrum of the modeled AGN, especially in the high-luminous objects. Surprisingly, when we tried to quantify the AGN fraction ($f_{agn}$), we found that the most luminous objects in the high-$z$ bin have a quite low $f_{agn} \sim 0.7-0.8$. This can further suggest the hypothesis of type II sources. In Appendix \ref{sec:Cigale_pdf}, we analyze the best-fit PDFs to check whether the second-best fit would change the type II to type I. We find that the fit for those objects is quite poor, as can be seen in Fig. \ref{fig:PDFs}, where the PDF for the $\Delta$ parameter is shown.
The PDF is flat, which results in a poor constraint on the $\Delta$ value.

\begin{table*}[]
\centering
\caption{No polar-dust: Median values of ratios of different estimations of luminosities for objects with L$_{disk}^{OPT}>10^{45}$ [erg/s]. The errors are the MAD errors. }

\begin{tabular}{|c|c|c|c|c|}
\hline
Column & Low & Intermediate 1  & Intermediate 2  & High \\
\hline
(1)L$^{torus}_{IR}$/L$^{photo}_{IR}$ & 0.74 $\pm$ 0.04 & 0.77 $\pm$ 0.07 & 0.80 $\pm$ 0.06 & 0.78 $\pm$ 0.09 \\
(2) L$_{IR}^{\text{torus+cold}}$/L$_{IR}^\text{{photo}}$ & 0.77 $\pm$ 0.04 & 0.79 $\pm$ 0.07 & 0.82 $\pm$ 0.06 & 0.80 $\pm$ 0.09 \\
(3) L$_{IR}^{\text{torus+cold+stellar}}$/L$_{IR}^\text{{photo}}$ & 0.83 $\pm$ 0.05 & 0.84 $\pm$ 0.04 & 0.83 $\pm$ 0.13 & 0.84 $\pm$ 0.09 \\
(4) L$_{IR}^{\text{torus+cold+stellar+disk}}$/L$_{IR}^\text{{photo}}$ & 1.00 $\pm$ 0.00 & 1.00 $\pm$ 0.00 & 1.00 $\pm$ 0.00 & 1.00 $\pm$ 0.00 \\
(5)L$_{OPT}^{disk}$/L$^{photo}_{OPT}$ & 0.92 $\pm$ 0.06 & 0.95 $\pm$ 0.07 & 0.94 $\pm$ 0.05 & 0.83 $\pm$ 0.14 \\
(6) L$_{OPT}^{disk+stellar\ attenuated}$/L$^{photo}_{OPT}$ & 1.00 $\pm$ 0.00 & 1.00 $\pm$ 0.00 & 1.00 $\pm$ 0.00 & 1.00 $\pm$ 0.01 \\
(7) L$_{OPT}^{disk+stellar\ attenuated}$/L$_{IR}^{disk}$ & 1.08 $\pm$ 0.08 & 1.05 $\pm$ 0.11 & 1.06 $\pm$ 0.08 & 1.21 $\pm$ 0.24 \\

\hline
\end{tabular}
\label{tab:No_polar_above}

\end{table*}

\begin{table*}[]
\centering
\caption{No polar-dust: Median values of ratios of different estimations of luminosities for objects with L$_{disk}^{OPT}<10^{45}$ [erg/s]. The errors are the MAD errors.}

\begin{tabular}{|c|c|c|c|c|}
\hline
Column & Low & Intermediate 1  & Intermediate 2  & High \\
\hline
(1)L$^{torus}_{IR}$/L$^{photo}_{IR}$ & 0.68 $\pm$ 0.11 & 0.75 $\pm$ 0.13 & 0.85 $\pm$ 0.05 & 0.86 $\pm$ 0.07 \\
(2) L$_{IR}^{\text{torus+cold}}$/L$_{IR}^\text{{photo}}$ & 0.70 $\pm$ 0.12 & 0.78 $\pm$ 0.13 & 0.88 $\pm$ 0.05 & 0.90 $\pm$ 0.06 \\
(3) L$_{IR}^{\text{torus+cold+stellar}}$/L$_{IR}^\text{{photo}}$ & 0.79 $\pm$ 0.13 & 0.90 $\pm$ 0.13 & 0.94 $\pm$ 0.04 & 0.99 $\pm$ 0.02 \\
(4) L$_{IR}^{\text{torus+cold+stellar+disk}}$/L$_{IR}^\text{{photo}}$ & 1.00 $\pm$ 0.00 & 1.00 $\pm$ 0.00 & 1.00 $\pm$ 0.00 & 1.00 $\pm$ 0.00 \\
(5) L$_{OPT}^{disk}$/L$^{photo}_{OPT}$ & 0.71 $\pm$ 0.40 & 0.87 $\pm$ 0.33 & 0.22 $\pm$ 0.34 & 0.02 $\pm$ 0.16 \\
(6) L$_{AGN}^{disk+stellar\ attenuated}$/L$^{photo}_{OPT}$ & 0.98 $\pm$ 0.04 & 1.00 $\pm$ 0.03 & 0.98 $\pm$ 0.03 & 0.97 $\pm$ 0.02 \\
(7) L$_{AGN}^{disk+stellar\ attenuated}$/L$_{IR}^{disk}$ & 1.38 $\pm$ 19.97 & 1.15 $\pm$ 6.07 & 4.43 $\pm$ 13.82 & 114.30 $\pm$ 2347.41 \\
\hline
\end{tabular}
\label{tab:No_polar_below}

\end{table*}

\subsection{Luminosity transformation}
\label{sec:transformation}
The main purpose of the luminosity cut at the threshold of L$^{disk}_{OPT}\sim10^{45}$ [erg/s] was to separate type I objects from outliers with significant contamination. Another goal was to check whether there is a simple transformation that changes the distribution of the photometric luminosity estimations to a distribution similar to the SKIRTOR L$_{disk}^{OPT}$ and L$_{torus}^{IR}$ estimations. 
This check should show whether the apparent photometry distribution can be easily calibrated to follow the AGN component's luminosity. We used the following method:

\begin{enumerate}
    \item Fit the ordinary least squares regression to the L$_{\text{photo}}$versus L$_{\text{CIGALE}}$ for both L$_{IR}$ and L$_{OPT}$ in the form $L^{X}_{CIGALE} = a \times L^{X}_{\text{photo}} + b$, where X is either IR or OPT.

    \item Use calculated coefficients to the  L$_{TRANSFORM}$ as follows: $L^{X}_{photo-tranformed} = a \times L^{X}_{\text{photo}} + b$.
    
\end{enumerate}

\begin{figure*}[!htp]

  \includegraphics[clip,width=1\textwidth]{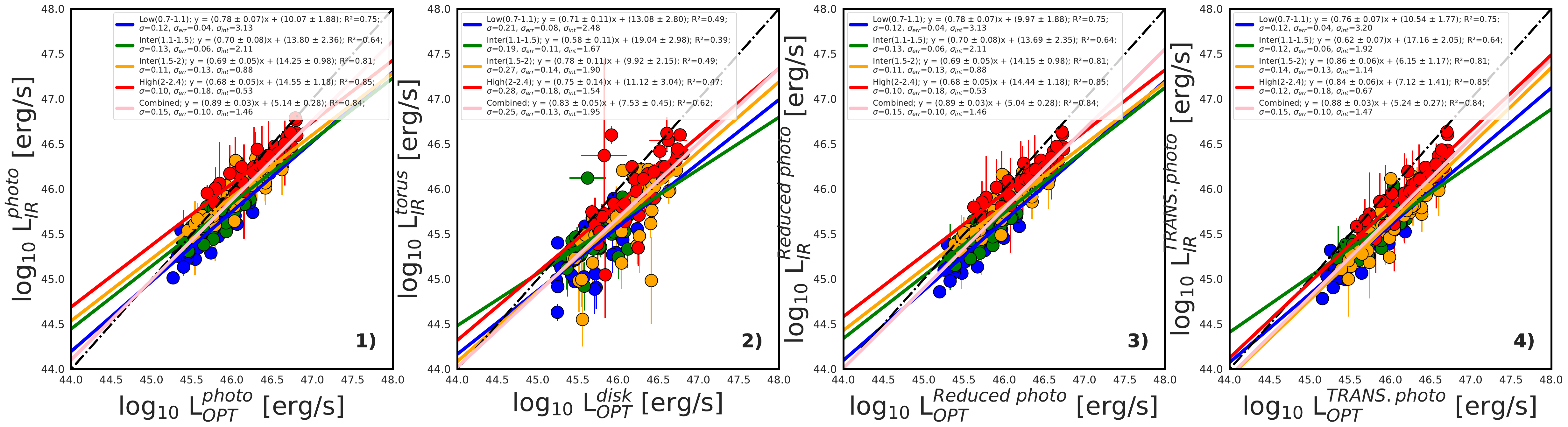}
  
\caption[width=0.5\columnwidth]{The transformation of luminosities/universal relation across redshifts for objects with L$_{OPT}^{disk}>10^{45}$ [erg/s]. The no. 1) plot shows the relation between photometric method estimated luminosities. Middle panel no. 2) shows the OPT-IR relation for CIGALE L$_{disk}^{OPT}$ and L$_{torus}^{IR}$, the second middle panel 3) shows the L$^{reduced\ photo}_{OPT}$ and L$^{reduced\ photo}_{IR}$ with reduced contaminations, described in the section \ref{sec:cont_reduced}. The right panel 4) shows the transformed luminosities, as described in section \ref{sec:transformation}. }
\label{fig:Lum_Tranformation}
\end{figure*}

\begin{figure*}[!htp]

  \includegraphics[clip,width=1\textwidth]{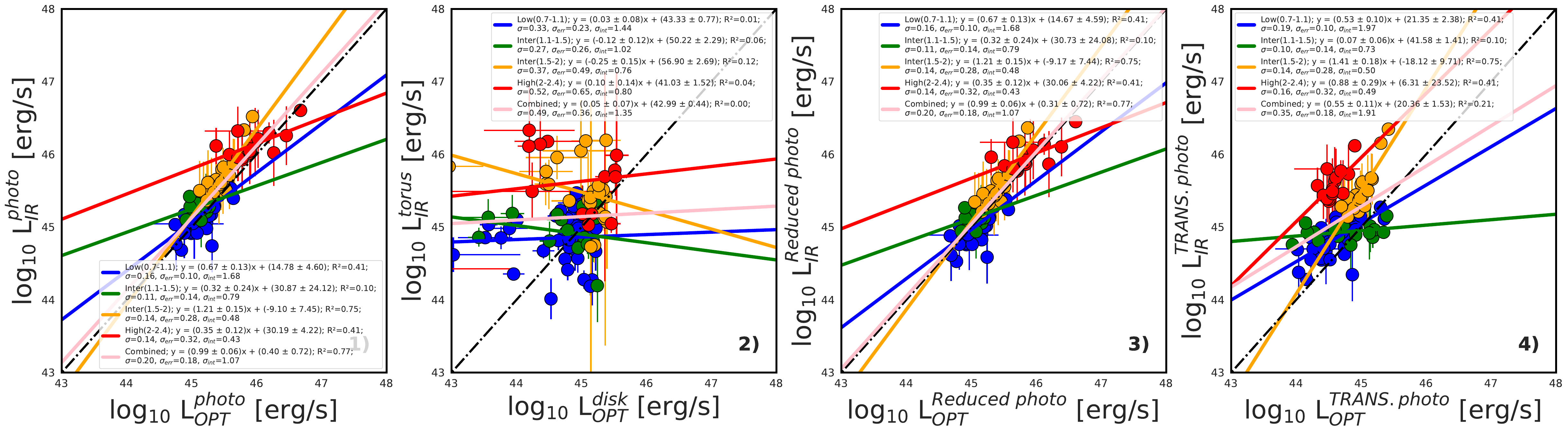}
  
\caption[width=0.5\columnwidth]{The transformation of luminosities/universal relation across redshifts for objects with L$_{OPT}^{disk}<10^{45}$ [erg/s]. For description see description under Fig. \ref{fig:Lum_Tranformation}.}
\label{fig:Lum_Tranformation_Dim}
\end{figure*}

The results of the luminosity transformation can be seen in Figs. \ref{fig:Lum_Tranformation} and \ref{fig:Lum_Tranformation_Dim} (for objects with L$_{disk}^{OPT} > 10^{45}$ [erg/s] and L$_{disk}^{OPT} < 10^{45}$ [erg/s] respectively). The transformation shifts the whole distribution toward lower luminosities, which correspond to the luminosities after extraction of the major contaminations from both photometric luminosities. The change is visible in the increase of the slopes for all redshift bins. The slopes are closer to those fit to the SKIRTOR estimations. The mean luminosities for all redshift bins are also closer to those of SKIRTOR. It can be seen that the transformation based on the CIGALE-estimated luminosities does not work for objects below L$_{disk}^{OPT} < 10^{45}$ [erg/s].

\subsection{Reduction of contaminations}
\label{sec:cont_reduced}
An alternative approach to the luminosity transformation proposed in Sect. \ref{sec:transformation} is to reduce the L$^{photo}_{IR}$ and L$^{photo}_{OPT}$ by a certain percentage. Based on our analysis and results shown in Table \ref{tab:comp_high}, we decided to reduce the luminosities for objects with L$^{disk}_{OPT}$>10$^{45}$ [erg/s] as follows: 1) L$^{Reduced/ photo}_{IR}$ = 0.7 $\times$ L$^{photo}_{IR}$ and 2) L$^{Reduced/ photo}_{OPT}$ = 0.85 $\times$ L$^{photo}_{OPT}$. The results of this approach can be seen in Fig. \ref{fig:Lum_Tranformation}, third panel. As one can see, the described approach seems to yield distributions that are similar to the transformed luminosities, but compared to those estimations, it does not change the slopes of the relation. An additional advantage of this method of luminosity transformation is that it is easy to reproduce with other data. The transformations have almost no effect on the overall scatter of the relation.

\section{Discussion}
\label{sec:Discussion}

Here, our goal is to calibrate a simple $L_{\text{OPT}}$–$L_{\text{IR}}$ relation based on more precise method of fitting the SED. By doing so, we can estimate the contributions of various components to the apparent photometric approach. As a result, it improves the accuracy of deriving AGN emission in the optical and IR bands from broadband photometry.
We calculated the intrinsic luminosities of key components, including stellar emission, the accretion disk, the dusty torus, polar dust, and cold dust. Subsequently, we evaluated their respective contributions within the typical photometric coverage in the optical and near-IR bands.

\begin{figure}[!htp]

\includegraphics[clip,width=1\columnwidth]{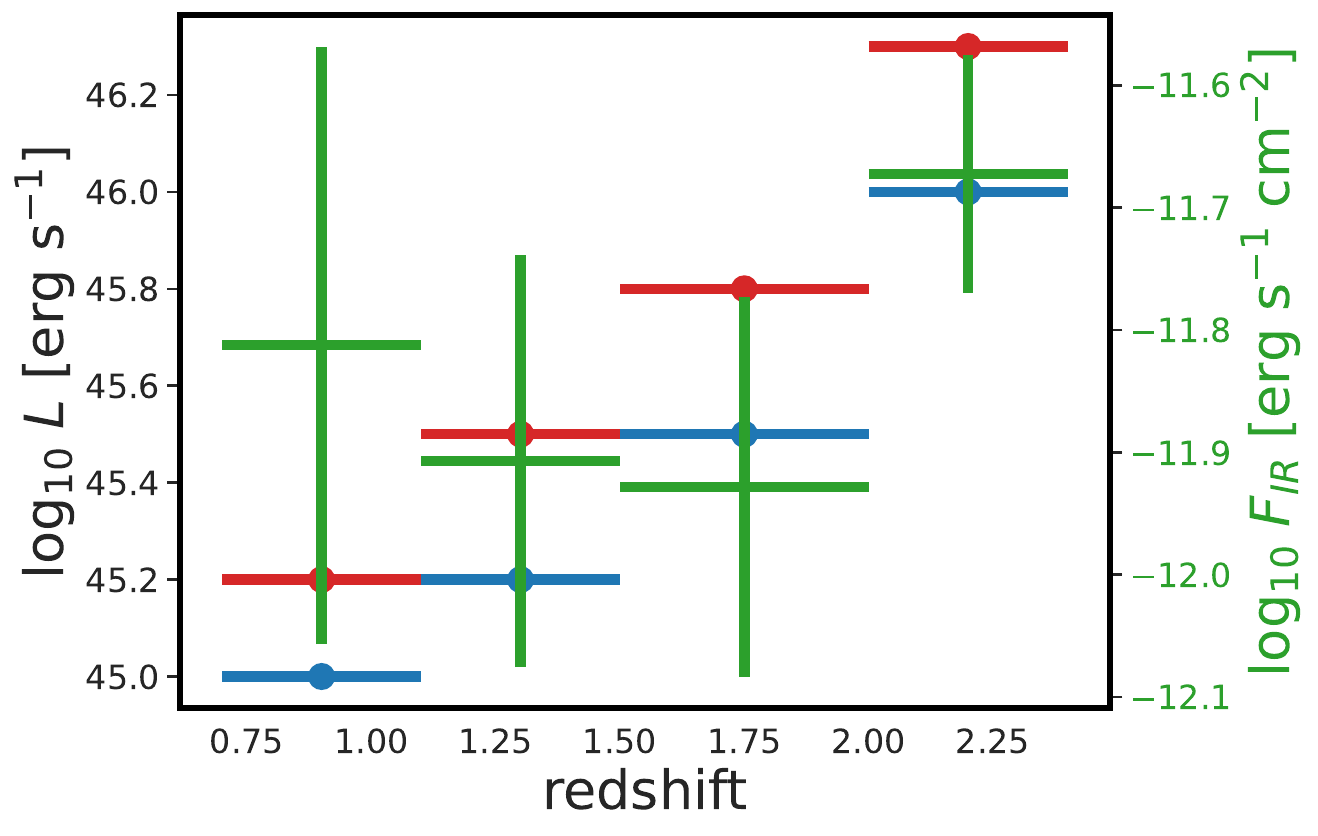}
\caption[width=0.5\columnwidth]{The transition luminosity above which the SED fit is dominated by the AGN components in the consecutive redshift bins. The red points shows L$_{IR}^{photo}$ which approximates the dusty torus and the green points reveal corresponding fluxes. The blue points visualize L$_{OPT}^{photo}$, which correspond to the accretion disk. }

\label{fig:Lum_Cut_Redshift}
\end{figure}

Our interest was in identifying the parameter range in which a simple estimation of the disk and torus luminosities is feasible. In particular, we investigated whether integrating the SED in two fixed wavelength intervals, 0.11--1 $\mu$m for optical (L$_{OPT}^{photo}$) and 1--7 $\mu$m for IR bands (L$_{IR}^{photo}$), can be reliably used, following \cite{Ralowski2024}. 
Figure~\ref{fig:Lum_Cut_Redshift} displays the threshold luminosities for L$_{OPT}^{photo}$ in blue and L$_{IR}^{photo}$ in red, below which the trend in correlation with the AGN components breaks in a given redshift bin. The approximate limits on optical log luminosity are 45, 45.2, 45.5, and 46 for the consecutive redshift bins, as shown in the first row of Fig.~\ref{fig:Luminosities_vs_contaminations}. Above these values, L$_{OPT}^{photo}$ strongly correlates with L$_{OPT}^{disk}$, and the accretion disk dominates the SED in the 0.11--1 $\mu$m range. For the IR, the respective log luminosity limits are 45.2, 45.5, 45.8, and 46.3, as visible in the fourth row of Fig. \ref{fig:Luminosities_vs_contaminations}. Those limits denote the lower end of the good behaving correlation of L$_{IR}^{photo}$ with L$_{IR}^{torus}$, so they give luminosities where the dusty torus dominates in the 1--7 $\mu$m band. Within these luminosity ranges, the contribution from the host galaxy components can be considered negligible, thus luminosity cut allows for robust photometric estimation of AGN emission. The final size of the sample is 171 objects.

The luminosity cut, which divides the sample into sources with a well-defined correlation between observed and modeled AGN luminosity and those with a broken correlation, can be translated into fluxes, as shown in Fig. \ref{fig:Lum_Cut_Redshift}. Our analysis indicated that these cuts in both the optical and IR bands are primarily driven by data quality. In the 0.11--1 $\mu$m range, the most important factor is the short-wavelength point, which in most cases is the NUV from GALEX (for low-$z$) and u from SDSS. This point determines the CIGALE solution, as it influences the classification between UV dominated by attenuated starlight (type 2 AGN) and accretion disk dominated UV (type 1 AGN). As a result, fainter sources often exhibit prominent outliers with the fit low disk luminosity. In the IR part, long wavelength points impose their limits. The W3 and W4 WISE filters are import for dusty torus and polar dust fit accuracy. The W4 is especially known to be flaky \citep[i.e.][]{Ralowski2024}. Additionally, Herschel data, especially for high-z sources, in the dimmer regime are challenging. For instance, their blending issue compromises the accuracy of constraining cold dust and polar dust contributions. Beyond observational issues, physical effects may also contribute to the dispersion. For example, in low-z less luminous sources are expected to be more affected by polar dust contamination \citep[i.e.][]{Toba_2021,Ralowski2024}. On the other side, z > 2 galaxies tend to exhibit higher dust content overall.

Based on Fig. \ref{fig:OA_i_PD}, we find that the majority of sources in the inclination versus opening angle plane are clustered within a triangular region. This distribution suggests that many objects fall into the intermediate category, where the edge of the dusty torus is expected to be close to the line of sight. Removing the polar dust component increases the opening angle by at least 20 degrees, while the inclination typically decreases.

The primary motivation for incorporating a polar dust component into the SED fitting procedure was to enhance the precision of addressing photometric points between 20 and 300 $\mu$m. In a subset of sources, the MIR and FIR points were insufficient. We find that polar dust can strongly influence the estimated physical properties of quasars. Although its luminosity is typically four orders of magnitude lower than that of the torus, the polar dust constitutes an important contaminant when estimating the parameters of the torus. In the presence of polar dust, the opening angle is on average smaller compared to the fit without it. This leads to a model-dependent degeneracy: either a solution with polar dust and a low opening angle or one without polar dust and a higher opening angle. This finding is similar to \cite{Toba_2021}, where the PD was also shown to influence the estimated covering factor, and the authors stated that the use of PD improves fit quality, based on the Bayesian information criterion. However, the authors of that study did not account for contamination from the ISM, specifically cold dust. When fitting the IR photometry up to W4, the cold dust influence in the longer wavelength starts to be significant (see Fig.  \ref{fig:sed_example}). It is therefore expected that including a model component dedicated to FIR emission (e.g., polar dust) leads to improved fits compared to models that exclude it. Nevertheless, even in our models without polar dust, the cold dust component alone does not fully reproduce the observed emission. \cite{Ciesla_2015} reported that the AGN contribution estimates with CIGALE are often overestimated for type 1 AGN and underestimated for type 2 AGN, with the threshold being an AGN fraction of around 50\%.

It has been reported that the SKIRTOR model does not produce deep enough silicate absorption features to fit the SEDs of ultraluminous IR galaxies \citep{Efstathiou_2022}. In CIGALE, slight modifications to the SKIRTOR model were implemented, in particular, the inclusion of the polar dust component. However, this addition appears insufficient to resolve the issue, as the polar dust component does not contribute significantly in the wavelength range where silicate features are prominent. This limitation was evident when comparing fits with polar dust and without polar dust.

However, the photometric method of luminosity estimation is not perfect, as it contains 15-30\% contaminations (L$_{IR}^{photo}$ and L$_{OPT}^{photo}$ respectively). Despite these uncertainties, the transformed luminosities agree well with SKIRTOR-based estimates of L$_{IR}^{torus}$ and L$_{OPT}^{disk}$. It should be noted that although the transformed luminosity values are close to SKIRTOR, they may still inherit the same model-dependent limitations discussed earlier. However, the transformed OPT-IR relation, visible in Fig. \ref{fig:Lum_Tranformation}, remains robust across the sample, with a similar spread before and after the transformations, around 0.15 for joined redshifts 0.7-2.4. The slope of the transformed OPT-IR relation for all redshift bins is in the range of 0.8-0.9, which suggests its potential applicability in cosmological studies \citep{QSO2025}. Analogous to the X-UV relation described in the introduction of this paper.

\section{Conclusions}
\label{sec:Conclusions}

In the presented work, the OPT-IR relation based on the large sample of OPT-UV quasars was analyzed. The major points we have found:

\begin{itemize}
    \item The major contaminations for L$_{IR}^{photo}$ were disk emission and cold dust, while polar dust have a minor influence. For L$_{OPT}^{photo}$ the major contamination was stellar emission.
    \item The accretion disk, starlight, and cold dust respectively contribute up to 17\%, 15\%, and 12\% to the L$_{IR}^{photo}$.
    \item The photometric luminosity estimation method works well for objects with a luminosity above L$^{disk}_{OPT}\sim$10$^{45}$ [erg/s] compared to the luminosity estimations from CIGALE.
    \item Instead of the cut on L$^{disk}_{OPT}$ we can minimize the influence of contaminations on optical and IR luminosity by using redshift-dependent cuts on L$_{OPT}^{photo}$ and L$_{IR}^{photo}$.
    \item The population of type 1 AGNs can be contaminated by misclassified sources of type 2 --- L$_{\rm disk}$ outliers. This misclassifications can result in the drop in accuracy of the SED fitting below the luminosity threshold. This effect can be strongly suppressed by imposing luminosity cuts.
    \item The proposed transformation for photometric luminosities produces the OPT-IR relation with a distribution similar to L$_{IR}^{torus}$, L$_{OPT}^{disk}$ estimated with the SKIRTOR model.
    \item The inclusion of the polar dust model changes the estimation of the physical properties of the quasars, such as the opening angle, inclination, and AGN fraction. This can be seen in the comparison between the fit with polar dust and the fit without polar dust (see Fig.~\ref{fig:with_vs_out}). However, it should be noted that the effect on luminosities is minor.
    \item The polar dust, which can be thought of as the extended atmosphere of the torus, is important in the lower luminosity end of the population (especially low-z, where it contributes up to 10\% L$_{\rm IR}$).
    \item Accounting for the polar dust improves fit quality and affects the inclination and the torus thickness estimations.
    
\end{itemize}

We believe that we identified the major contaminations within the L$_{OPT}^{photo}$ and L$_{IR}^{photo}$. The calibration procedure for the OPT-IR relation is crucial to use it in the future as cosmological probe.

 \begin{acknowledgements}
 
 MR has been supported by the Polish National Agency for Academic Exchange (Bekker grant BPN/BEK/2024/1/00298). This research was supported by the Polish National Science Center grants 2018/30/M/ST9/00757 and 2023/50/A/ST9/00579 and by Polish Ministry of Science and Higher Education grant DIR/WK/2018/12. This research was funded by the Priority Research Area Digiworld under the program Excellence Initiative - Research University at the Jagiellonian University in Kraków. We gratefully acknowledge Polish high-performance computing infrastructure PLGrid (HPC Center: ACK Cyfronet AGH) for providing computer facilities and support within computational grant no. PLG/2024/017491. This article is based upon work from COST Action CA21136 – “Addressing observational tensions in cosmology with systematics and fundamental physics (CosmoVerse)”, supported by COST (European Cooperation in Science and Technology). This research has made use of the NASA/IPAC Infrared Science Archive, which is funded by the National Aeronautics and Space Administration and operated by the California Institute of Technology. {\it Herschel} is an ESA space observatory with science instruments provided by European-led Principal Investigator consortia and with important participation from NASA.
 \end{acknowledgements}

\bibliographystyle{aa-note} 
\bibliography{ref} 

\begin{appendix}

\section{Testing Cigale PDFs}
\label{sec:Cigale_pdf}

\begin{figure*}[!htp]

  \includegraphics[clip,width=1\textwidth]{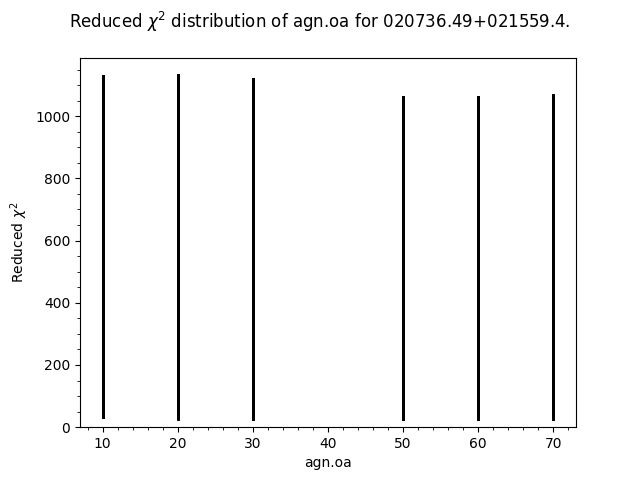}
  
\caption[width=0.25\columnwidth]{An example PDF plot for $\Delta$ parameter for one of the problematic galaxies at high-$z$.}
\label{fig:PDFs}
\end{figure*}

After running an additional test for outliers after the luminosity cut described in section \ref{sec:Lum_Cut}, we considered whether this group of objects is wrongly fitted by the Cigale. Our hypothesis was, that some of those objects were wrongly classified as Type II source, and the solution for Type I had only slightly smaller probability distribution function (PDF) for values of parameters described as type I. We performed additional SED fitting to calculate PDFs for SKIRTOR and stellar models. The example PDFs are shown in figure \ref{fig:PDFs}. As on can see the $\chi^2$ values are extremely similar for this object, and it is unclear whether the chosen value 60 [deg] is truly the best one.

\begin{figure*}[!htp]

  \includegraphics[clip,width=1\textwidth]{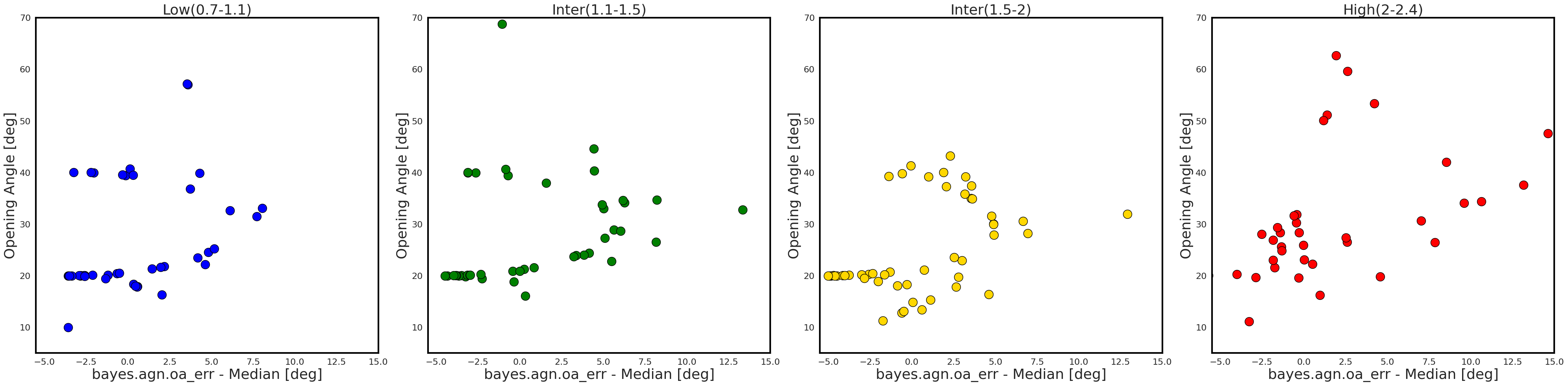}
  
\caption[width=0.5\columnwidth]{The $\Delta$ error analysis. On Y axis the $\Delta$ is presented. On X axis the difference between $\Delta _{err}$ and median of $\Delta _{err}$ is shown. It is visible that 1) for high-$z$ the general error is higher, 2) the errors increase, when CIGALE estimates are further from given a priory values.

}
\label{fig:Median_oa_er_full}
\end{figure*}

\section{OPT-IR relation vs f$_{AGN}$}
\begin{figure*}[!htp]

  \includegraphics[clip,width=1\textwidth]{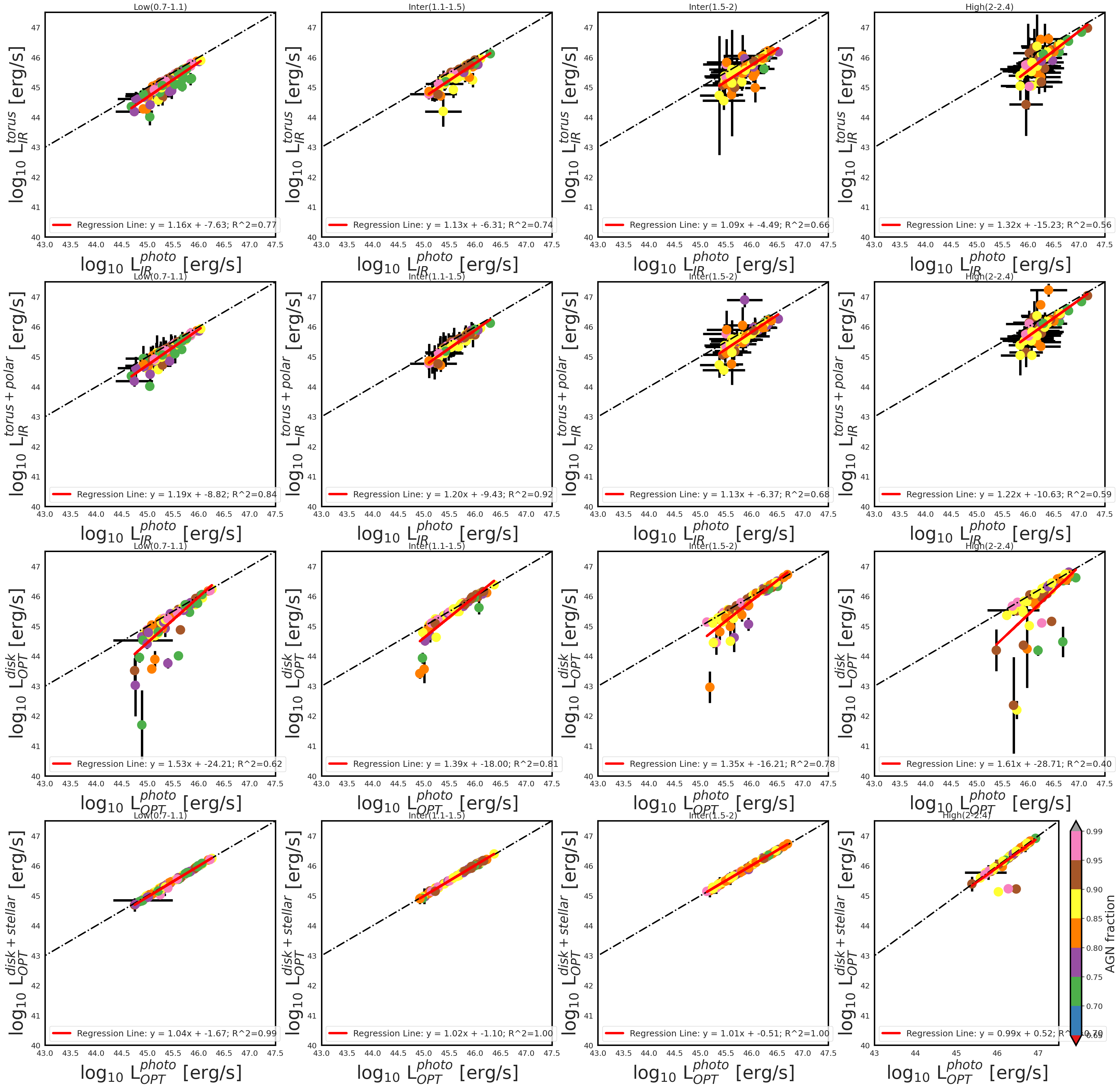}
  
\caption[width=1\textwidth]{Comparison of luminosities estimated from 1) our photometric method (presented on X-axis) and 2) CIGALE estimates without and with contaminations for 4 redshift bins. }
\label{fig:Luminosities_vs_contaminations_appendix}
\end{figure*}

\begin{figure*}[!htp]

  \includegraphics[clip,width=1\textwidth]{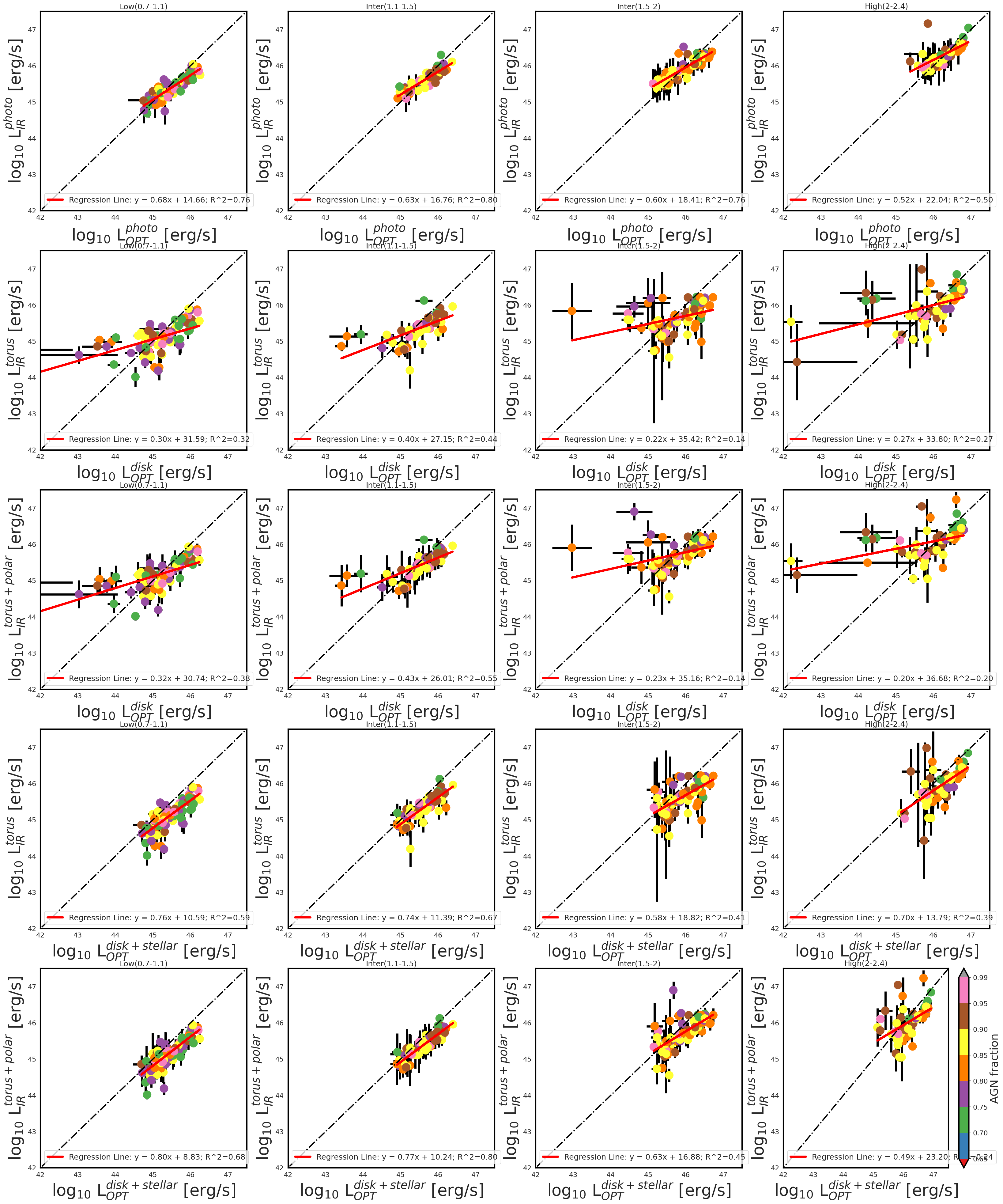}
  
\caption[width=0.5\columnwidth]{The OPT-IR relation for all redshift bin through different estimates of luminosities. In the first row the $L_{IR}$ and $L_{OPT}$ are estimated with photometric method. For the relation to be robust the cut in the luminosities may be necessary. }
\label{fig:Comparison_OPT-IR_appendix}
\end{figure*}

Figures \ref{fig:Luminosities_vs_contaminations_appendix} and \ref{fig:Comparison_OPT-IR_appendix} are modifications of figures \ref{fig:Luminosities_vs_contaminations} and \ref{fig:Comparison_OPT-IR}, the color representing the f$_{AGN}$. Figure \ref{fig:Luminosities_vs_contaminations_appendix} shows that the relation with f$_{AGN}$ is not obvious. On one side the low values of f$_{AGN}$ are present with low luminosity outliers, see 3rd row, which would mean that those objects could be recognized as type II AGN. On the other hand, few of high luminous objects also have low f$_{AGN} < 0.8$.

\section{SFR vs M$_{*}$}
\begin{figure*}[!htp]

\caption{SFR vs M$_{*}$ for our sample of quasars. }
  \includegraphics[clip,width=1\textwidth]{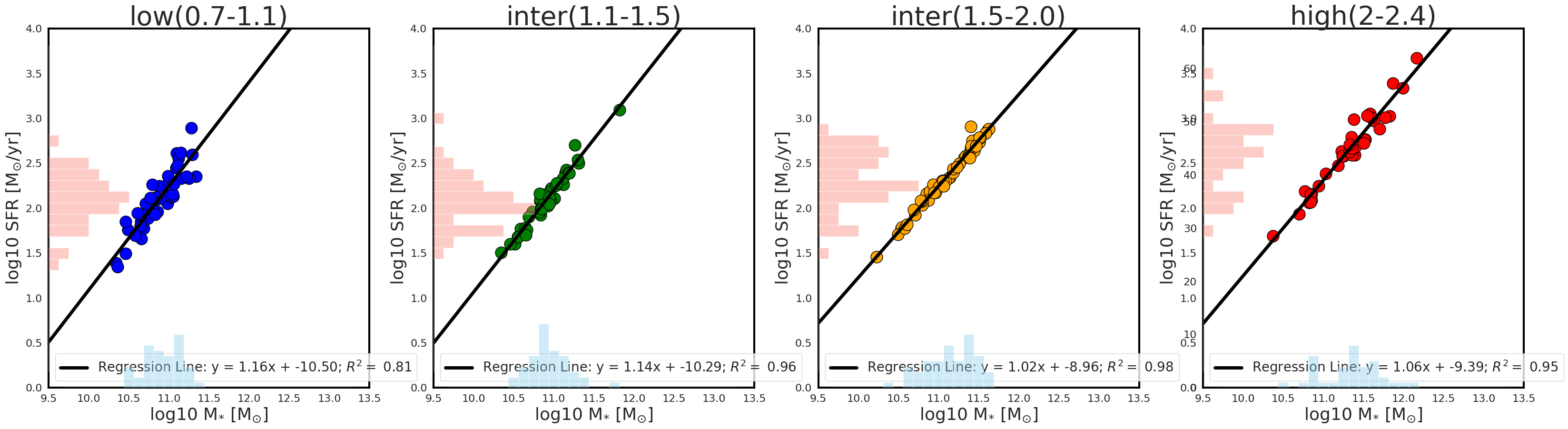}

\label{fig:SFRvsM*}
\end{figure*}

We checked the estimated stellar masses (M$_{*}$) and Star Formation Rate (SFR) on Fig. \ref{fig:SFRvsM*}. In recent work by \cite{Buat2021} the authors also analyzed SED for quasars with CIGALE. We wanted to compare the received distributions of M$_{*}$ and SFR to the results therein. \cite{Buat2021} calculated the SED with, and without polar dust component, and then compare SFR and M$_{*}$ histograms (see Fig. 9 therein). Distributions for our samples in figure \ref{fig:SFRvsM*} resemble values calculated with polar dust.

\section{SED example}
\begin{figure*}[!htp]

  \includegraphics[clip,width=1\textwidth]{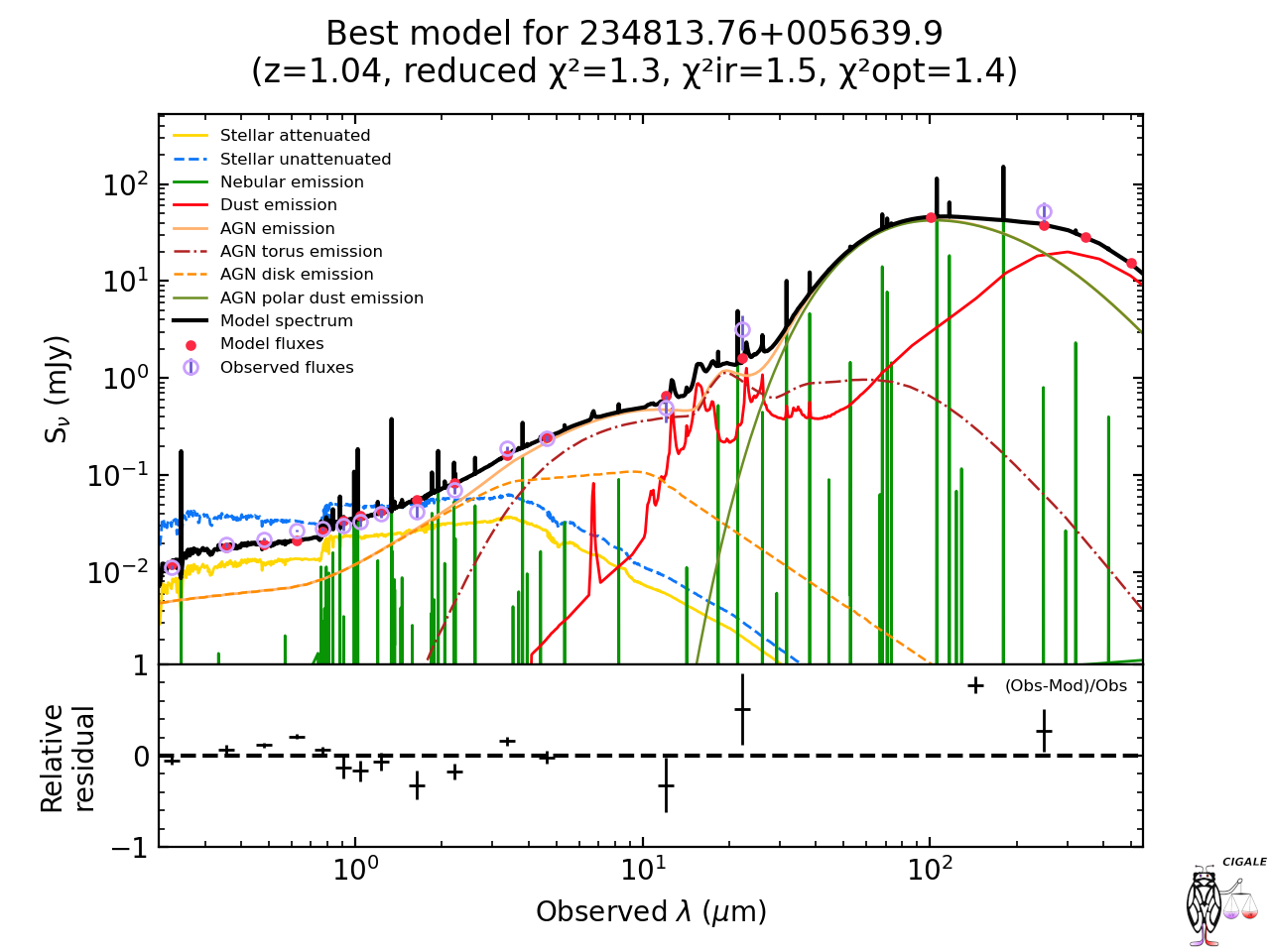}
  
\caption[width=1\columnwidth]{An example SED for one of quasars in observer frame. The upper panel shows the best-fitted SED model, which is showed with thick, black line. The components of model spectrum are stellar emission \citep{BC2003} (both attenuated - solid, yellow line, and unattenuated - dashed blue line), AGN emission with orange color \citep{Skirtor_2016}, cold dust emission with red, solid line \citep{Dale2014}. The red dots mark the modeled flux observations, based on the broadband filters transmissions. The pink circles mark the photometrical observations in the observed-frame of the objects. The border of integration between OPT and IR luminosities is at 1$\mu$m, in rest-frame and 2.04$\mu$m in observer frame.}
\label{fig:sed_example}
\end{figure*}

Below we present the SED plot generated with CIGALE code in Figure \ref{fig:sed_example}. The plot was slightly modified to calculate the reduced $\chi^2$ in IR and OPT domains. The black, solid line represents the model spectrum which is the sum of stellar, AGN, cold dust models.

\section{Diagnostic plots}
\begin{figure}[!htp]

  \includegraphics[clip,width=1\textwidth]{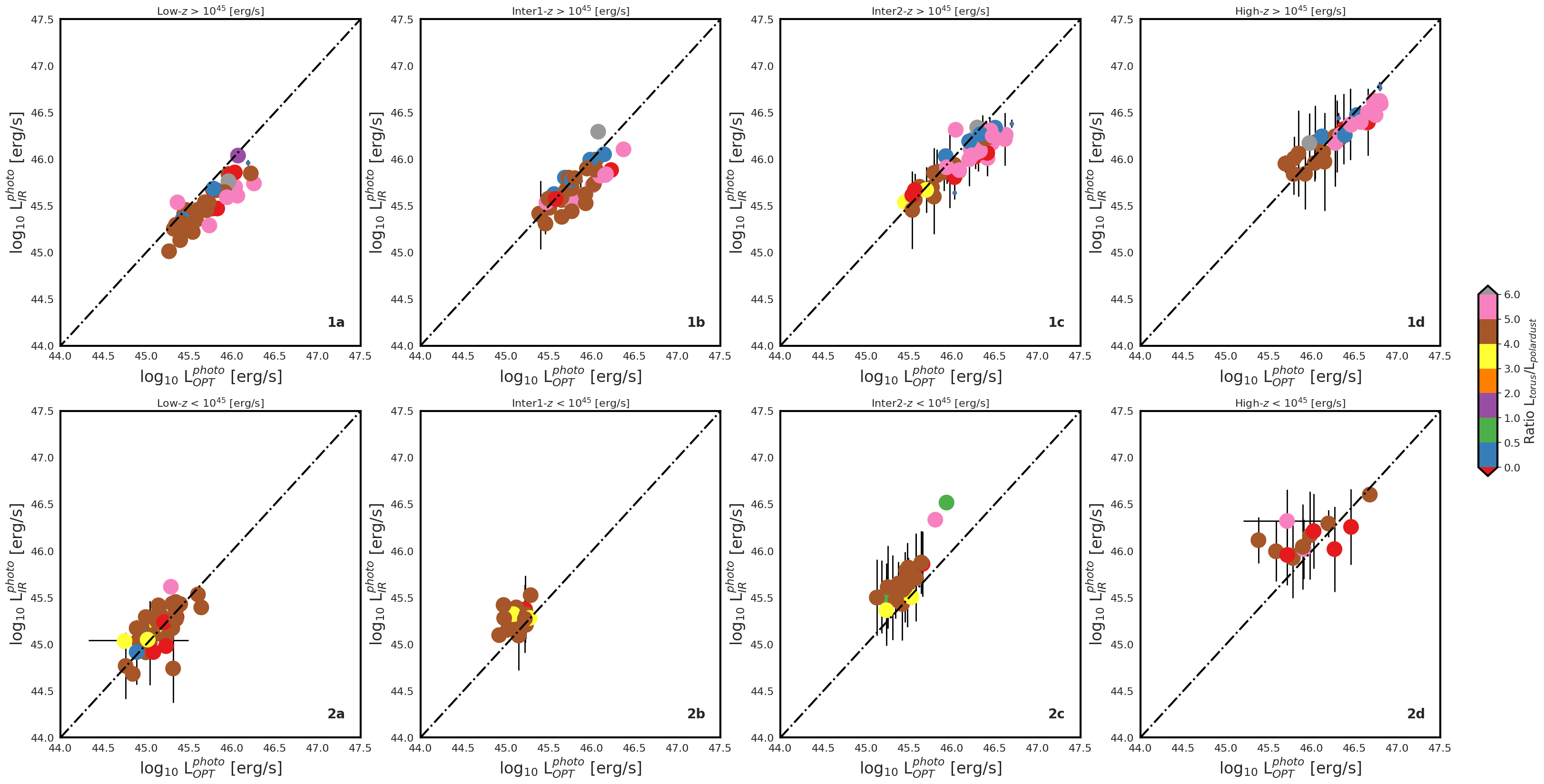}

\caption[width=1\columnwidth]{The L$_{IR}$ vs L$_{OPT}$ relation for all data samples. Upper panel shows the objects with L$^{disk}_{OPT} > 10^{45}$ [erg/s], while lower shows dimmed objects with L$^{disk}_{OPT} < 10^{45}$ [erg/s]. With the color the ratio of L$_{IR}^{polar dust}$/L$_{IR}^{torus}$ is shown.}
\label{fig:all_ratio}
\end{figure}

\begin{figure}[!htp]

  \includegraphics[clip,width=1\textwidth]{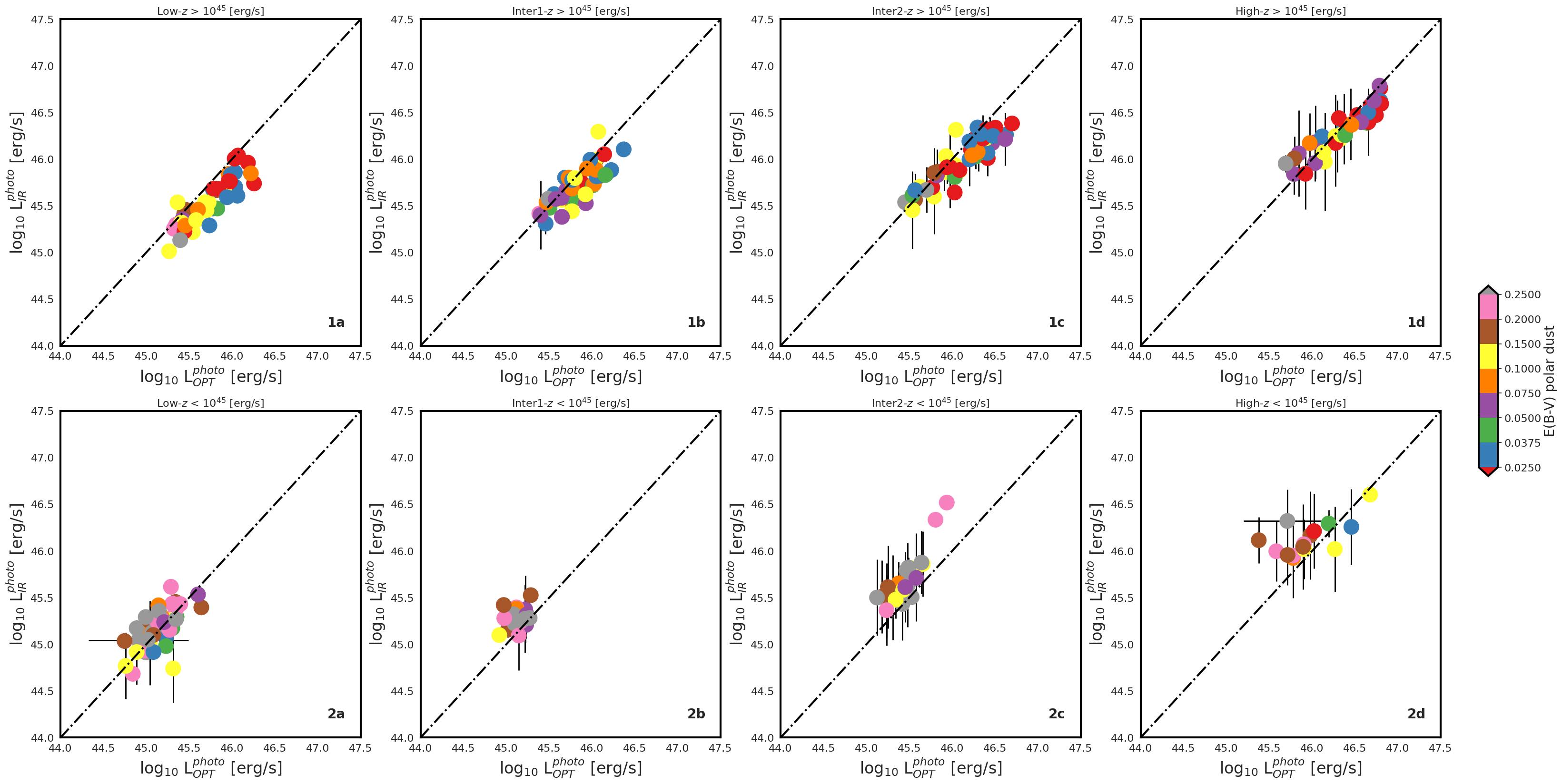}

\caption[width=1\columnwidth]{The L$_{IR}$ vs L$_{OPT}$ relation for all data samples. Upper panel shows the objects with L$^{disk}_{OPT} > 10^{45}$ [erg/s], while lower shows dimmed objects with L$^{disk}_{OPT} < 10^{45}$ [erg/s]. With the color the E(B-V) for polar dust is shown.}
\label{fig:All_E(B-V)}
\end{figure}

The additional diagnostic plots for the L$_{OPT}^{photo}$ vs L$_{IR}^{photo}$ relation, showing with the color the E(B-V)$_{PD}$ in the figure \ref{fig:All_E(B-V)} and the ratio of L$_{torus}$/L$_{PD}$ visible with color on Figure \ref{fig:all_ratio}. The Figure \ref{fig:All_E(B-V)} show that a number of objects below the luminosity threshold have the highest values in E(B-V)$_{PD}$. There are still some pbjects with the lowest luminosities above the threshold with similarly high E(B-V)$_{PD}$, which suggests that not all of the contaminated objects were excluded with the luminosity cut, but for the majority of objects the luminosity cut work well, as on average the objects with lower values of E(B-V)$_{PD}$ are left.

In Figure \ref{fig:all_ratio} shows that the objects above the luminosity threshold on average have torus luminosity 5 or more times higher than the PD luminosity. The objects below the threshold have lower ratio, though still few times higher than PD luminosity.

\newpage

\begin{figure*}[!htp]

  \includegraphics[clip,width=1\columnwidth]{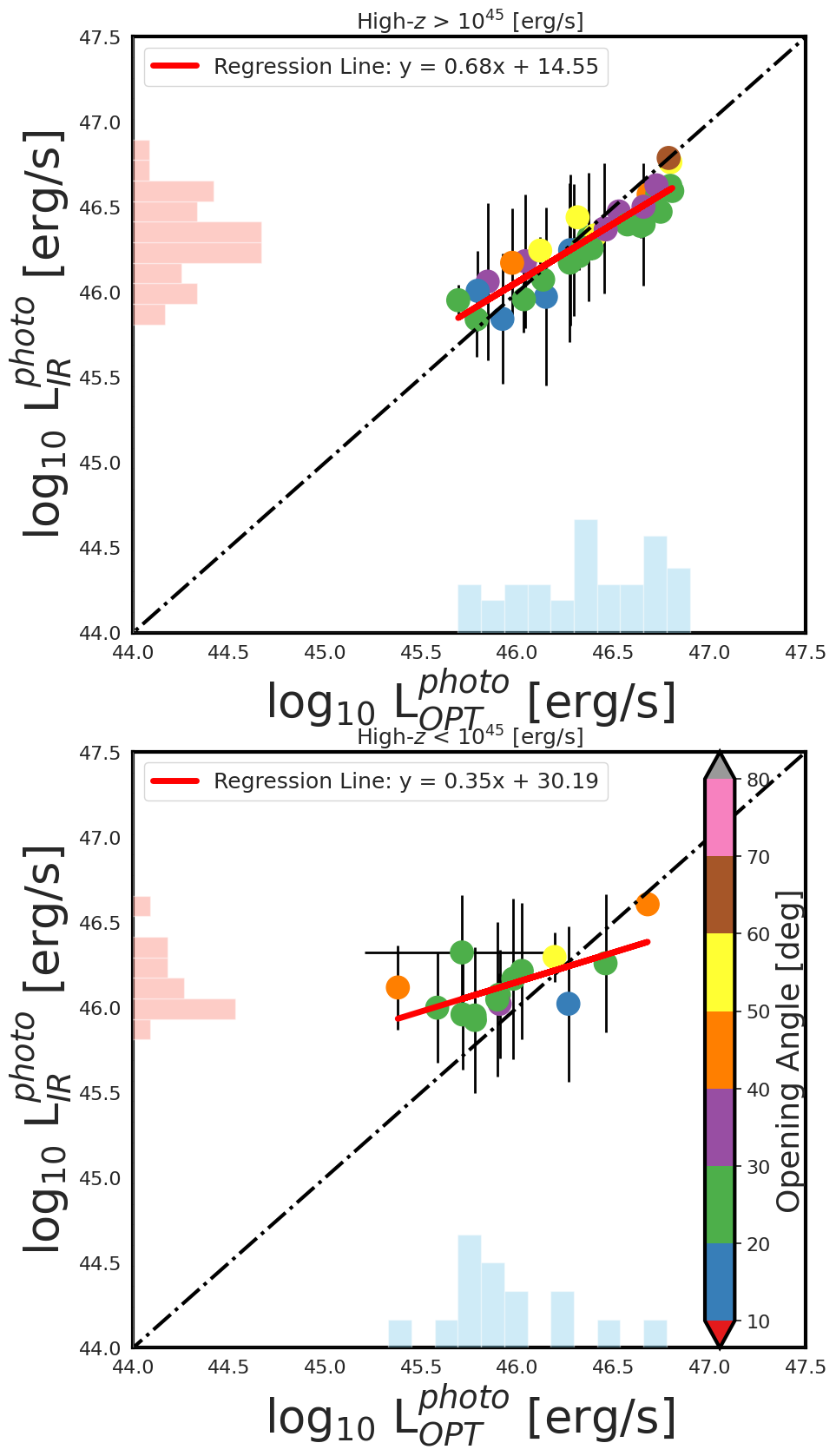}

\caption[width=1\columnwidth]{The L$_{IR}$ vs L$_{OPT}$ relation for high-$z$ sample. Upper panel shows the objects with L$^{disk}_{OPT} > 10^{45}$ [erg/s], while lower shows dimmed objects with L$^{disk}_{OPT} < 10^{45}$ [erg/s]. With the color the half opening angle of the dusty torus is shown.}
\label{fig:high_OA}
\end{figure*}

\begin{figure*}[!htp]

  \includegraphics[clip,width=1\columnwidth]{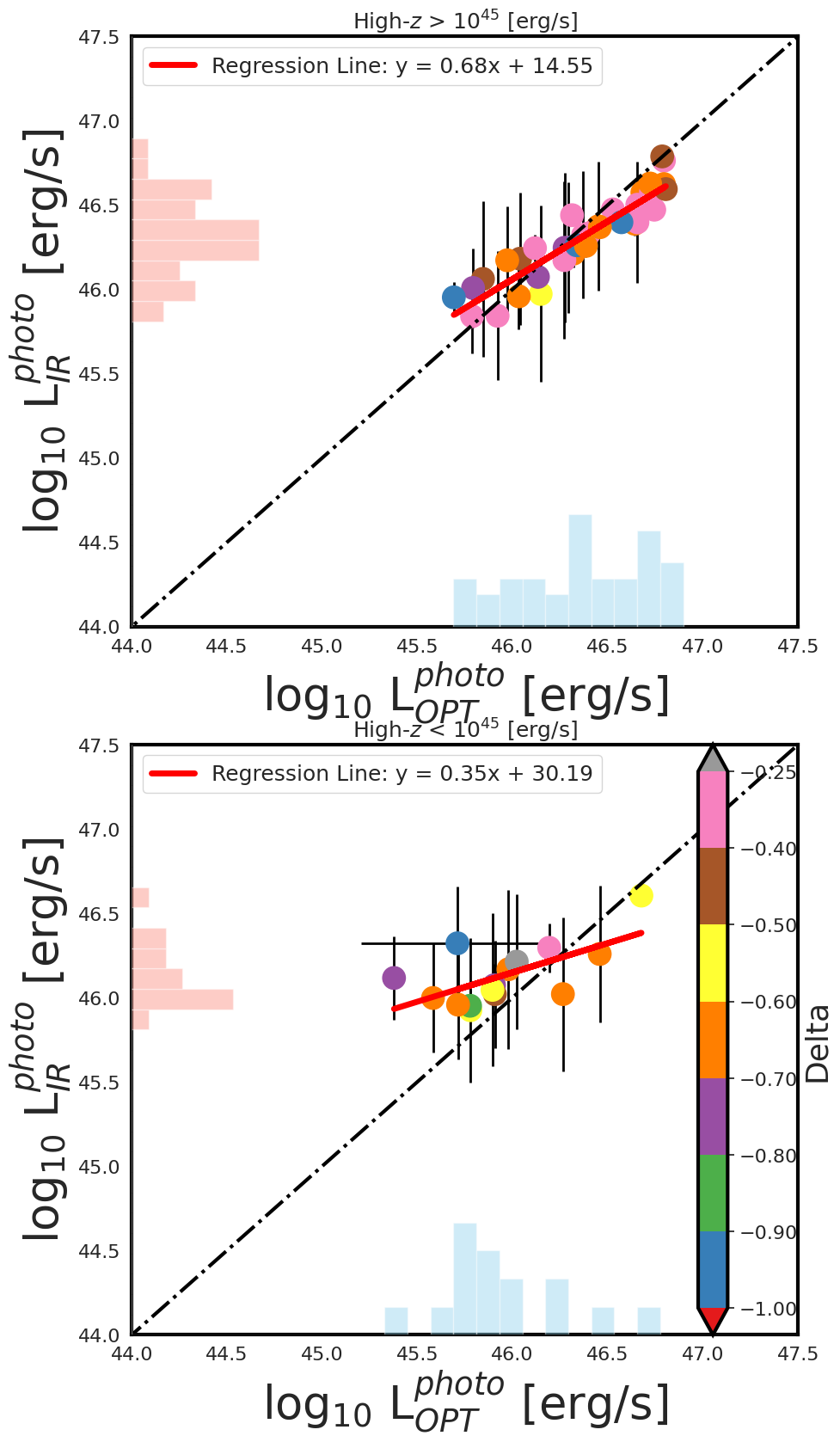}

\caption[width=1\columnwidth]{The L$_{IR}$ vs L$_{OPT}$ relation for high-$z$ sample. Upper panel shows the objects with L$^{disk}_{OPT} > 10^{45}$ [erg/s], while lower shows dimmed objects with L$^{disk}_{OPT} < 10^{45}$ [erg/s]. With the color the 'delta' slope of SKIRTOR is shown.}
\label{fig:high_delta}
\end{figure*}

\end{appendix}

\end{document}